\input harvmac
\def\R{{\bf R}}
\def\Q{{\bf Q}}

\def\K3{{\bf K3}}
\def\journal#1&#2(#3){\unskip, \sl #1\ \bf #2 \rm(19#3) }
\def\andjournal#1&#2(#3){\sl #1~\bf #2 \rm (19#3) }

\def\hat{\widehat}

\def\tilde{\widetilde}

\def\frac#1#2{{#1\over#2}}

\def\half{\frac12}

\def\d{\partial}

\def\inbar{\,\vrule height1.5ex width.4pt depth0pt}
\def\IC{\relax\hbox{$\inbar\kern-.3em{\rm C}$}}
\def\IR{\relax{\rm I\kern-.18em R}}
\def\IP{\relax{\rm I\kern-.18em P}}
\def\Z{{\bf Z}}

%
%
\def\np#1(#2)#3{Nucl. Phys. {\bf B#1} (#2) #3}
\def\pl#1(#2)#3{Phys. Lett. {\bf #1B} (#2) #3}
\def\plb#1(#2)#3{Phys. Lett. {\bf #1B} (#2) #3}
\def\prl#1(#2)#3{Phys. Rev. Lett. {\bf #1} (#2) #3}
\def\physrev#1(#2)#3{Phys. Rev. {\bf D#1} (#2) #3}
\def\prd#1(#2)#3{Phys. Rev. {\bf D#1} (#2) #3}
\def\ap#1(#2)#3{Ann. Phys. {\bf #1} (#2) #3}
\def\prep#1(#2)#3{Phys. Rep. {\bf #1} (#2) #3}
\def\rmp#1(#2)#3{Rev. Mod. Phys. {\bf #1} (#2) #3}
\def\cmp#1(#2)#3{Comm. Math. Phys. {\bf #1} (#2) #3}
\def\cqg#1(#2)#3{Class. Quant. Grav. {\bf #1} (#2) #3}
\def\mpl#1(#2)#3{Mod. Phys. Lett. {\bf #1} (#2) #3}
\def\atmp#1(#2)#3{Adv. Theor. Math. Phys. {\bf #1} (#2) #3}
\def\jhep#1(#2)#3{J. High Energy Phys. {\bf #1} (#2) #3}

\def\math#1{math.AG/#1}
\def\qalg#1{q-alg/#1}
\def\alggeom#1{alg-geom/#1}
\def\am#1(#2)#3{Ann. Math. {\bf #1} (#2) #3}
\def\im#1(#2)#3{Invent. Math. {\bf #1} (#2) #3}

\catcode`\@=11
\def\slash#1{\mathord{\mathpalette\c@ncel{#1}}}
\overfullrule=0pt
\def\AA{{\cal A}}
\def\BB{{\cal B}}
\def\CC{{\cal C}}
\def\DD{{\cal D}}

\def\FF{{\cal F}}
\def\GG{{\cal G}}

\def\LL{{\cal L}}
\def\NN{{\cal N}}
\def\MM{{\cal M}}
\def\OO{{\cal O}}

\def\SS{{\cal S}}

\def\lam{\lambda}

\def\underrel#1\over#2{\mathrel{\mathop{\kern\z@#1}\limits_{#2}}}

\catcode`\@=12


%

\def\det{{\rm det}}

\def\tr{{\rm tr}}
\def\Tr{{\rm Tr}}
\def\rk{{\rm rk}}

\def\sinh{{\rm sinh}}

\def\det{{\rm det}}
\def\exp{{\rm exp}}
\def\ch{{\rm ch}}
\def\chern{{\rm c}}
\def\End{{\rm End}}
\def\NS{{\rm NS}}
\def\Pic{{\rm Pic}}
\def\Ker{{\rm Ker}}
\def\Tors{{\rm Tors}}
\def\Hom{{\rm Hom}}


\def\zbar{{\bar z}}

\def\psibar{{\bar\psi}}
\def\phibar{{\bar\phi}}

\def\gammabar{{\bar\gamma}}

\def\dbar{{\bar \d}}

\def\ferm{{\vartheta}}


\lref\bfss{T. Banks, W. Fischler, S. Shenker and L. Susskind, {\it M Theory
           as a Matrix Model: A Conjecture}, \physrev55(1997)5112,
           hep-th/9610043.}
\lref\exclusion{J. Maldacena and A. Strominger, {\it $AdS_3$ Black Holes
                and a Stringy Exclusion Principle},
                \jhep9812(1998)005, hep-th/9804085.}
\lref\largen{J. Maldacena, {\it The Large N Limit of Superconformal Field
                Theories and Supergravity},
                \atmp2(1998)231, hep-th/9711200.}
\lref\seibergwitten{N. Seiberg and E. Witten, {\it The $D1/D5$ System and
                Singular CFT}, \jhep9904(1999)017, hep-th/9903224.}
\lref\macentropy{A. Strominger, {\it Macroscopic Entropy of $\NN=2$ Extremal
                Black Holes}, \plb383(1996)39, hep-th/9602111.}
\lref\micentropy{A. Strominger and C. Vafa, {\it Microscopic Origin of The
                Bekenstein-Hawking Entropy}, \plb379(1996)99, hep-th/9601029.}
\lref\commads{A. Giveon, D. Kutasov and N. Seiberg, {\it Comments on String
                Theory on $AdS_3$}, \atmp2(1998)733, hep-th/9806194.}   
\lref\entropywit{J. Maldacena, A. Strominger and E. Witten, {\it Black Hole
                Entropy in M-Theory}, \jhep9712(1997)002, hep-th/9711053.}
\lref\symmbreak{I. Klebanov and E. Witten, {\it AdS/CFT Correspondence and
                Symmetry Breaking},
                \np556(1999)89, hep-th/9905104.}
\lref\lehn{M. Lehn, {\it Chern Classes of Tautological Sheaves on Hilbert
             Schemes of Points on Surfaces}, \im136(1999)157, \math9803091.}
\lref\cysingular{I. Klebanov and E. Witten, {\it Superconformal Field Theory
                 on Threebranes at a Calabi-Yau Singularity},
                 \np536(1998)199, hep-th/9807080.}
\lref\cftcy{S. Gukov, C. Vafa and E. Witten, {\it CFT's from Calabi-Yau
                 Four-folds}, hep-th/9906070.}
\lref\vafa{C. Vafa, {\it Black Holes and Calabi-Yau Threefolds},
                 \atmp2(1998)207, hep-th/9711067.}
\lref\baryandbrane{E. Witten, {\it Baryons and Branes in Anti de Sitter
                  Space},
                  \jhep9807(1998)006, hep-th/9805112.}
\lref\adswitten{E. Witten, {\it Anti de Sitter Space and Holography},
                  \atmp2(1998)253, hep-th/9802150.}  
\lref\dijkgraaf{R. Dijkgraaf, {\it Instanton Strings and Hyperk\"ahler
                Geometry}, \np543(1999)545, hep-th/9810210.}
\lref\dmvv{R. Dijkgraaf, G. Moore, E. Verlinde and H. Verlinde, {\it
                   Elliptic Genera of Symmetric Products and Second
                    Quantized String}, \cmp185(1997)197, hep-th/9608096.}
\lref\dvv{R. Dijkgraaf, E. Verlinde and H. Verlinde, {\it Matrix String
                  Theory}, \np500(1997)43, hep-th/9703030.}
\lref\harvmoore{J. Harvey and G. Moore, {\it On the Algebra of BPS States},
                   \cmp197(1998)489, hep-th/9609017.}
\lref\zaslovyau{S.-T. Zaslov and E. Yau, {\it BPS States, String Duality and
                   Nodal Curves on K3},
                   \np471(1996)503, hep-th/9512121.}
\lref\nakajimaa{H. Nakajima, {\it Heisenberg Algebra and Hilbert Schemes of
                   Points on Projective Surfaces},
                   \am145(1997)379, \math9507012.}
\lref\nakajimab{H. Nakajima, {\it Lectures on Hilbert Schemes of Points on
                Surfaces}, University Lecture Series {\bf 18}, Providence, RI:
                American Mathematical Society. xi, (1999),  page 132.}
\lref\nakajimac{H. Nakajima, {\it Jack Polynomials and Hilbert Schemes of Points
                on Surfaces}, \alggeom9610021.}
\lref\vafapuzz{C. Vafa, {\it Puzzles at Large N}, hep-th/9804172.}
\lref\deBoera{J. de Boer, {\it Six-Dimensional Supergravity on 
              $S^3\times AdS_3$ and 2d Conformal Field Theory},
               \np548(1999)139, hep-th/9806104.}
\lref\deBoerb{J. de Boer, {\it Large N Elliptic Genus and $AdS/CFT$
              Correspondence}, \jhep9905(1999)017, hep-th/9812240.}
\lref\smallinst{E. Witten, {\it Small Instantons in String Theory}, 
            \np460(1996)541, hep-th/9511030.}
\lref\strong{C. Vafa and E. Witten, {\it A Strong Coupling Test of S-Duality},
            \np431(1994)3, hep-th/9408074.}
\lref\douglas{I. Brunner, M. Douglas, A. Lawrence, C. R\"omelsberger, {\it
              D-branes on the Quintic}, hep-th/9906200.}
\lref\ktheorya{E. Witten, {\it D-Branes and K-Theory}, \jhep9812(1998)019,
              hep-th/9810188.}
\lref\ktheoryb{G. Moore and E. Witten, {\it Self-Duality, Ramond-Ramond
               Fields and K-Theory}, \jhep0005(2000)032, hep-th/9912279.}
\lref\ktheoryc{D.-E. Diaconescu, G. Moore and E. Witten, {\it E8 Gauge Theory, 
               and a Derivation of K-Theory from M-Theory}, hep-th/0005090.}
\lref\modulid{M. Douglas and G. Moore, {\it D-branes, Quivers and ALE
              Instantons}, hep-th/9603167.}
\lref\bsv{M. Bershadsky, V. Sadov and C. Vafa, {\it D-Branes and Topological
              Field Theories}, \np463(1996)420, hep-th/9511222.} 
\lref\hananywitten{A. Hanany and E. Witten, {\it Type IIB Superstrings, BPS
              Monopoles, and Three-Dimensional Gauge Dynamics},
              \np492(1997)152, hep-th/9611230.}
\lref\solwitten{E. Witten, {\it Solutions of Four-Dimensional Field Theories
               via M Theory}, \np500(1997)3, hep-th/9703166.}
\lref\wittenhiggs{E. Witten, {\it On the Conformal Field Theory of the Higgs
              Branch}, \jhep9707(1997)003  hep-th/9707093.}
\lref\johnson{C. Johnson, {\it More Superstrings from Supergravity}, 
              \np537(1999)129, hep-th/9805047.}
\lref\viele{E. Bergshoeff, B. de Wit and M. Vasiliev, {\it The Structure of
          the Super-$W_{\infty}(\lambda)$ Algebra}, \np366(1991)315.}
\lref\class{E. Witten, {\it Overview of K-Theory Applied to Strings},
             hep-th/0007175.}
\lref\guenadin{M. G\"unaydin, G. Sierra and P. Townsend, {\it The Unitary
           Supermultiplets of $D=3$ Anti de Sitter and $D=2$ Conformal
           Superalgebras}, \np274(1986)429.}
\lref\vinet{L. Lapointe and L. Vinet, {\it Exact operator solution of the
           Calogero-Sutherland model}, \cmp178(1995)435, \qalg9509003.}
\lref\pope{C. Pope, {Lectures on W algebras and W gravity}, hep-th/9112076.}
\lref\li{M. Li, {\it Strings from IIB Matrices}, \np499(1997)149,
           hep-th/9612222.} 
\lref\ellipgenus{E. Witten, {\it Elliptic Genera and Quantum Field Theory}, 
                \cmp109(1987)525.}
\lref\hull{C. Hull, {\it Duality and the Signature of Space-Time}, 
          \jhep9811(1998)017, hep-th/9807127.}
\lref\neks{N. Nekrasov and A. Schwarz, {\it Instantons on Noncommutative
           $R^4$, and $(2,0)$ Superconformal Six Dimensional Theory}, 
	   \cmp198(1998)689, hep-th/9802068.}
\lref\ftheory{C. Vafa, {\it Evidence for F-Theory}, \np469(1996)403, hep-th/9602022.}



\Title{
\rightline{hep-th/0011292}} 
{\vbox{\centerline{W Algebras from AdS/CFT Correspondence}}}
\medskip

\centerline{\it Stephan Korden \foot{korden@physik.rwth-aachen.de}}
\bigskip
\centerline{Institute for Theoretical Physics, RWTH Aachen}
\centerline{D-52056 Aachen, Germany}

\smallskip

\vglue .3cm
\bigskip
\noindent
We consider a system of $D5/D1$ branes in the supergravity background
$AdS_3\times S^3\times X$, where $X$ is $T^4$ or $K3$. By investigating the
structure of the missing states in the conformal description, we are able to
extend the $AdS/CFT$ correspondence to W algebras. As a test of this new
formulation the results are compared to Hilbert schemes and more general
supergravity backgrounds as deformations by D3-branes or six-dimensional
Calabi-Yau manifolds.
\noindent
\Date{11/2000}


\newsec{Introduction}

The introduction of branes in addition to the classical fundamental string
brought two new aspects into the low energy formulation of the different
string sectors. One is the anomaly term of D-branes and its
understanding in the context of K-theory. This way, M theory provides the
RR-charges with a natural intersection form and thus a lattice structure
in K-theory \ktheoryb. A seemingly different ansatz pursues matrix
theory. For each string sector there exists a matrix or ``little string''
theory, which is build up from an infinite set of branes in the infinite
momentum frame of M theory. Following the $AdS/CFT$ conjecture \largen,
these six-dimensional low energy limits are related to  
$AdS_3\times S^3\times X$ supergravity backgrounds \refs{\bfss,\johnson},
where $X$ is  either $T^4$ or $K3$. In this article we will mainly address
the second point and investigate the supergravity theory of 
$AdS_3\times S^3\times K3$. 

In many aspects the $AdS/CFT$ duality in three dimensions is special. At
first gravity in the $AdS_3$ background is a topological theory with a
two-dimensional CFT on its boundary. Therefore, all dynamical
degrees of freedom have to originate from the compact part of spacetime. The
purely topological nature of the theory makes quantisation practicable
\refs{\exclusion, \commads, \seibergwitten}, and it is the only example
showing that the KK-spectrum and the modes of the
conformal field theory coincide. It has been Vafa \vafapuzz, who first
observed that the chiral primaries and its
descendants of the CFT are not enough to account for all states in the KK
spectrum. Later, this puzzle has been solved by observing that some of the
multiparticle states from supergravity correspond to non-chiral primaries,
which itself are not descendants from any other chiral fields but the
product of descendants of chiral primaries \refs{\deBoera, \deBoerb}. And
although the missing states can be constructed this way, they are not part
of the original CFT spectrum as would have been expected by the $AdS/CFT$ duality.
To solve this problem we propose a correspondence between $AdS$ and a 
supersymmetric $W_{\infty}(\lambda)$ algebra.

To get a better understanding how W algebras enter the discussion, we will
summarise the main ideas as follows. Take for example $Q_5$ D5-branes and $Q_1$ 
D1-branes wrapped on $X$, leaving a string in the remaining six dimensions.
As long as no further fields are turned on, the branes can freely join and separate.
The corresponding effective 1+1 dimensional field theory of this system
contains a $SU(Q_5)$ gauge group with $Q_1$ additional instantons. The two
extremal constellations, with all branes separated or joined,
translate to the Coulomb respectively Higgs branch of the low energy theory. This
passage between two sectors with all its intermediate states are the
``dynamical'' degrees of freedom. But moving branes in a curved
background are not well understood because of lack of a good description for
the low energy sector. This is different for the moduli
space $\MM_{(Q_5,Q_1)}$ which is expected to be equivalent to a $(4,4)$
sigma model on the target space $\rm{Sym}^{N}(X)$ for $N=Q_5Q_1$
\refs{\seibergwitten, \bsv}. But this description has two major drawbacks. First,
$N$ has to be large and $Q_5$ and $Q_1$ have to be relatively prime, otherwise the moduli
space is reducible and the representation as a symmetric product is not valid,
and second, the moduli space depends on the product of the charges only,
which contradicts the effective field theory description. Both
problems have a solution by forming the infinite sum
\eqn\coproduct{ \MM = 
\sum_{N=1}^{\infty}\;\coprod_{\nu:\;\rm{partition\;of\;N}}[X]^{\nu}\;.}
Similar to the calculation of the Poincar\'{e} polynomials of Hilbert
schemes by G\"{o}ttsche's formula, the infinite sum over all partitions of
$N$ leads to important simplifications. We will show that, if the cohomology
classes of $X$ can be represented by a chiral ring in the topological
SCFT, the cohomology classes of $\MM$ are accessible by the
corresponding supersymmetric $W_{\infty}(1/4)$ algebra.

How does this construction solve the problem of the missing states?
The first nontrivial
example is the distribution of three D1-branes in an infinite stack of
D5-branes, each of charge one. Because only the product $Q_5Q_1$ of
charges enters the construction, it is always possible to embed a finite
system of D1-branes into an infinite stack of D5-branes. This way there are only three
different partitions for $N=3$
\eqn\part{ [3,0,0,\ldots]\;,\hskip 2em [2,1,0,\ldots]\hskip 1em \rm{and}\;
           \hskip 1em [1,1,1,\ldots]\;.}
The first and the third vector resemble primary states of the original
SCFT and are related by T duality. They correspond to the Higgs
respectively the Coulomb branch of the low energy description and have been
studied in \refs{\commads, \seibergwitten}. The coproduct in \coproduct\ reduces
to a simple product and the moduli space factorises which leaves a
Liouville theory as SCFT. The second partition in \part, however, represents one
of the missing states. In analogy to representation theory, the
previous problem of classifying the physical states on the CFT side,
splits into two parts. The first uses representation theory of the
symmetric group, as seen above. Whereas the second part relies on the topology
of $X$ only and not on that of its symmetric products. This assertion needs further
explanation.

Following the previous example, consider the primary state corresponding to
the partition $[1,1,1,\ldots]$.
Now it is a classical result from rational conformal field theory, that its 
null states correspond to Schur polynomials which, on the other hand, are
representations of the symmetric group. These special polynomials can be generated
by repeated operation with a differential operator on the lower lying
partitions. With reference to the above example, the state $[2,1,0,\ldots]$ 
is the result of this operator acting on
$[1,1,0,\ldots]$. Of course, there is no difference whether one is calculating
the null state from the Liouville theory and then operating with the shift
operator or one is first acting with the operator on the Liouville Lagrangian and
then constructing the null state. As we will show, the second possibility
has a general solution in form of a $W_{\infty}(\lambda)$ algebra, which
will be the main result of this article. In this setting, the duality between
a supersymmetric $AdS$ background and a conformal field theory is completely
determined by the Liouville action which is one reason why the original
$AdS/CFT$ correspondence works so well. Making use of results from Hilbert schemes
of singular points \refs{\nakajimaa, \nakajimab, \nakajimac, \lehn}, we will show that 
the only topological information entering the conformal field theory are the
Euler characteristic and the canonical class of $X$. Especially the article
\lehn\ by M. Lehn has been very important for understanding the
mathematical aspects of the construction.

This paper is organised as follows. In Section 2, we review the conformal
field theory of the untwisted sector as introduced in \refs{\commads,
\seibergwitten} and explain the puzzle of the missing states for the twisted
sector \refs{\vafapuzz, \deBoera}. Section 3 deals with the distribution of
D1-branes on a stack of D5-branes and its representation as Young diagrams.
We compare these to the null states of the previous section and review the
relation to Schur, respectively Jack polynomials. The purpose of Section 4 is
to review the mathematical definition of moduli spaces and its
representation as schemes. From the lectures of Nakajima \nakajimaa\
we present two different formulations, one as a field theory and one in terms of a
Virasoro algebra. The W algebra is constructed in Section 5 and possible
generalisations are considered. The final Section 6 contains the conclusions
and further suggestions for future investigation.


\newsec{String Theory on $AdS_3\times S^3\times X$}

In this section we review the compactification of string theory in the
$AdS_{D+1}\times \NN$ background. Here $\NN$ will be a compact
manifold whose holonomy group is large enough for at least some
supersymmetry to survive. According to the work of Seiberg and Witten
\seibergwitten\ only the case $D=2$ is in general stable under quantisation.
In the following it will be therefore necessary to consider the
conformal field descriptions of $AdS_3$ along the lines of \refs{\exclusion,
\commads} as well as the reduction to the large brane \seibergwitten. The
first representation is appropriate for demonstrating the problem of the
missing states in the CFT description \refs{\vafapuzz, \deBoera, \deBoerb},
whereas the second version is necessary for the construction of the W
algebra in Section 5. 

\subsec{Classical Results}

In \largen\ Maldacena proposed a remarkable correspondence between a fixed
supergravity background $AdS_{D+1}\times \NN$ and a supersymmetric conformal
field theory depending on the dimension $D$ and $\NN$. The mapping between
these two descriptions includes a stack of $N$ black D-1 branes. From a more
topological point of view, these branes carry a gravitational instanton
\foot{Here and in the following we will understand a topological configuration
as an instanton.} of charge $N$. Basically there is no great
difference between a classical gravitational instanton and a gauge instanton
in Euclidean space. In both cases, the charge of the topological solution
acts as an additional coordinate in the moduli space, and although the
moduli space has no natural metric, it carries a conformal structure which
is often sufficient to determine its topology. The $AdS/CFT$ correspondence
now states that, in the limit $N\to\infty$, both moduli spaces coincide or
stated differently, their boundaries meet at this point. That it is exactly
one point, can been seen from the example stated in the introduction. It is
not important where the D1-branes are located on the D5-brane for the limit
$Q_5\to\infty$, which is the reason why the target space of the sigma model
depends on the product $Q_5Q_1$ only. Using the picture of Maldacena, the
topological properties of gravitational instantons are mapped to the boundary of
$AdS_{D+1}$ with additional Chern-Simons terms located at $N$ arbitrary points
on its boundary. 

The hope is that not only the two moduli spaces join each other along
the boundary, but that it is actually possible to move from one region to
the other in a definite way. A similar phenomenon is known from mirror symmetry, where the
K\"ahler and the complex structure guarantee for a meromorphic moduli space
in the large complex structure limit, but even in this case it is not
sure that the two moduli spaces intersect in more than one point. But, if one
accepts the idea that there is a description of quantised supergravity
in terms of gauge instantons, there has to be a mapping for all finite brane
configurations and not only in the limit $N\to\infty$. Because the moduli space of
gauge instantons is conformally invariant, we have to assume
the same on the side of supergravity. So, take a subset
$W\subset AdS_{D+1}$ with the restriction that the boundary $\d W$ carries
a conformal structure. This is exactly the case studied in \seibergwitten,
with $W$ large and $\d W$ near the boundary of $AdS_{D+1}$. Although it is 
not evident, that this is a valid description of the underlying moduli space, it
still reproduces the results as expected from the field theory approximation. 

In contrast to the microscopic formulation of the D-brane modes in the
$AdS/CFT$ correspondence the large brane can be analysed by classical
calculations. The first important information one gets in this limit is the
underlying structure of the moduli space and its stability, which depends
strongly on the dimension of the $AdS$ space and the signature of the scalar
curvature. As shown in \seibergwitten, the boundary of $AdS_{D+1}$ has a
natural conformal structure but no natural metric. Thus the classical form 
\eqn\adsmetric{ds^2=r_0^2 (dr^2+\sinh^2{r}\;d\Omega^2)\;}
only gives rise to a boundary with a metric of gauged conformal
group. Because this gauge fixes the scalar curvature at the boundary, it
eliminates the freedom to deform the theory to different boundaries and thus
to different string vacua. It is thus not possible 
to study the stability of the moduli space. The problem
can be solved by a simple reparametrisation of the radial coordinate.
Instead of the metric \adsmetric\ with its fixed spherical boundary, it is
easier to begin with an arbitrary metric $ds^2=g_{ij}dx^idx^j$ on $\d W$
of fixed conformal structure. This metric has an unique embedding into $W$ by 
\eqn\confmet{ds^2={r_0^2\over t^2}\left(dt^2+\hat{g}_{ij}(x,t)dx^idx^j\right)\;,}
with the boundary condition
\eqn\bla{\hat{g}_{ij}(x,0) = g_{ij}(x)\;.}
A Taylor expansion in the variable $t$ near the boundary relates the
conformal parameter to the radial coordinate of \adsmetric\ by $t=2e^{-r}$.
Standard $AdS/CFT$ calculations then show that the variable $t$ is related to the
physical scalar field $\phi$ by the scaling relation $\phi\sim t^{-(D-2)/2}$
for $D>2$ and a logarithmic dependence for $D=2$, which is typical for a
Liouville field. The final relation between the radial coordinate $r$ and
the physical relevant field $\phi$ can then be summarised by
\eqn\radial{r=\cases{{2 \over D-2}\log{\phi}+{1\over (D-1)(D-2)}\phi^{-{4\over D-2}}R
                                            & for $D>2$ \cr
		     \phi +e^{-2\phi}\phi R & for $D=2$}}
Obviously the two-dimensional case behaves quite differently and one could 
argue that this is only a pathological case. But later in this section we will
show how $\phi$ is related to the eleventh dimension in M theory, and thus 
plays a fundamental role in many brane interactions.

Now that the metric and its physical field content are known, the classical
Lagrangian of a BPS saturated large brane in $D$ dimensions can be calculated
from the DBI and WZW action \seibergwitten
\eqn\final{S =\cases{\frac{Tr_0^D}{2^{D-3}(D-2)^2}\int\sqrt g\left(
              (\partial\phi)^2+{D-2\over 4(D-1)}\phi^2 R+
	      \CO(\phi^{2(D-4)\over D-2}) \right) & for $D>2$;\cr
              {Tr_0^2\over 2}\int\sqrt g\left((\partial\phi)^2 +\phi R
	      -\half R  + \CO(e^{-2\phi }) \right)& for $D=2$.\cr}}
Because we started with an Euclidean version of $AdS$, the integration is
over the compact space $S^D$ respectively $\d W$. An analogous discussion
for the Minkowski space should be possible but its results for the moduli
space is not clear to us, so that we will omit this point.

What is the main difference between the two-dimensional boundary and the
case $D>2$? For $\phi$ large and constant the integrand of \final\ reduces to
the potential terms $\phi^2 R$ and $\phi R$. It is not very surprising that
the sign and thus the stability of a physical state of the dominant part of
the field theory description depends only on the scalar curvature of the
conformal space $\d W$. For $D>2$, the BPS saturated D-branes are free to
move relatively to each other as long as only one type of brane is involved.
This picture changes of course if additional branes are included. But we will
show that in the moduli space a sector of Liouville type develops, which is of the
same type as the one found for $D=2$. The basic information we get for the
moduli space of $D>2$ thus is that a monoculture of D-branes has a rather
trivial moduli space, and we believe that the $AdS/CFT$ correspondence is
actually exact. One example is the case of a stack of black D3-branes in an
$AdS_5\times S^5$ background. The dual conformal field theory is $\NN=4$
supersymmetric in four dimensions with $SU(N)$ gauge theory. The only
contribution to the otherwise trivial beta function are instanton
corrections. This is a strong evidence that the moduli space is basically the
moduli space of $SU(N)$ gauge instantons.

Things change dramatically in the case of $D=2$ or nontrivial brane
interactions. The $\phi R$ part in \final\ is not only a potential on
the D-brane action, but a topological term, proportional to the conformal charge 
in the Liouville action. Furthermore the moduli space develops an infinite
tube if a D1-brane instanton shrinks to zero size and separates from the 
D5-brane, dividing the Coulomb from the Higgs branch \wittenhiggs. But, if
this picture is correct, the opposite version has to be valid, too, namely
that the moduli space of any D-brane configuration can be constructed from
two basic elements, the Coulomb and the Higgs branch of one D1-instanton only.
This is exactly what we will do in Section 5.

\subsec{The Liouville Theory}

Here we will study the Higgs branch of the Liouville theory in the
background of one D1-brane. As stated above, the understanding of this sector
is the first step in constructing moduli spaces of intersecting branes.

Before entering the construction of the two-dimensional conformal field
theory, some details concerning the configuration of the branes are
necessary. The D5-branes are located in the $(x_0, x_1, x_6, x_7, x_8, x_9)$
plane of the ten-dimensional space, whereas the D1-branes are stretched in
the $x_0, x_1$ directions. The last four dimensions of the D5-brane are
wrapped on a manifold $X$, where $X$ is either $T^4$ or $K3$. Somewhere on
this compact space $X$ each D1-brane is fixed in one point, and to simplify our
discussion we will assume that all these points are located along the $x_6$
coordinate. To complete the brane spectrum for the type IIB string theory we
insert D3-branes in the $(x_0, x_1, x_8, x_9)$ plane. This choice of
coordinates suggests that the D3-branes are unaffected by the introduction
of the D1-branes which is definitely not true. But as a starting point it
simplifies the construction of the corresponding CFT considerably. And as
further motivation, one can imagine that the set of fixed points on $X$ are
free to move and thus can be located at the coordinates $x_6, x_7$. In
terms of the CFT on $X$, the dynamics of the D-branes decouple, but later on
the vanishing of the total conformal charge intertwine the degrees of
freedom and generates an interaction between the different types of branes.

In analogy to the considerations in \refs{\hananywitten, \solwitten}, the
D-brane configuration can be interpreted in terms of M theory. For this we
enlarge the ten-dimensional string description by the variable 
$x_{10}$, compactified on a sphere $S^1$ of radius $R$. The coordinate $t$
of the metric \confmet\ can then be identified as the complex moduli parameter 
\eqn\tmod{ t = 2\;\exp{(-(x_6+i x_{10})/R)}}
from the eleventh dimension, whereas the additional parameter of the D3-branes 
\eqn\vmod{ v = x_8 + i x_9 }
has no entry up to now. The moduli parameters $(t,v)$ specify the positions
of the branes as the roots of a polynomial $F(t,v)$. The interpretation of
$F$ as an algebraic curve and its connections to gauge instantons is well
understood \refs{\hananywitten, \solwitten}, but its connection to the
$AdS/CFT$ correspondence allows a more direct interpretation from the
viewpoint of M theory. Take for example the string limit $R\to 0$. From the
definition \tmod\ we see that $t$ vanishes and the metric \confmet\ reduces
to the boundary of $W$, while the number of D5-branes goes to infinity.
Therefore, the complex ``radius'' of $AdS$ has to be identified with 
$r=(x_6+i x_{10})/R$, which is proportional to $N=Q_5Q_1$. 
At first view the complex value of $r$ for
$x_6\neq 0$ may seem to contradict our identification, but actually it is
not $r$ we have to compare but the complex field $\phi$, or stated
differently, one needs at least one D1-brane to ensure a complex field.
What is the interpretation of the moduli parameter \vmod\ in the context of
$AdS/CFT$? The sources of D3-brane charges are orbifold
singularities on $W$. One explicit example for $AdS_5\times S^5$ with an
analysis of the moduli space can be found in \symmbreak. Thus $v$
parametrises the deformation of the conformal metric $\hat{g}_{ij}(x,t;v)$ 
in \confmet. Because the interpretation of D3-branes along the lines of \commads\ as an
orbifold on $\d W$ is not very intuitive. This is why we leave it to Section 5
to give a further analysis along the lines of Hilbert schemes. 

In the following we review the conformal field theory on 
$AdS_3\times S^3\times T^4$ near the boundary of $AdS_3$. In this limit the
metric \adsmetric\ reduces to 
\eqn\hilfa{ds^2 = dr^2 + e^{2r}\d\gamma\d\bar{\gamma}\;.}
After the analytic continuation of the radial coordinate, the worldsheet Lagrangian is 
\eqn\hilfb{\LL = \bar\d\phi\d\phi+e^{2\phi}\bar\d\gamma\d\bar\gamma\;,}
and can be put into the standard form of a Liouville action after
introducing an auxiliary field $\beta$ and the improvement term from \final
\eqn\liv{\LL = \bar\d\phi\d\phi-{2\over \alpha_+}\hat R\phi +
               \beta\bar\d\gamma+\bar\beta\d\bar\gamma -
	       \beta\bar\beta\;\exp{\left(-{2\over \alpha_+}\phi\right)}\;,}
with the Liouville parameter $\alpha_+^2=2k-4$. In the near horizon limit,
the conformal field theory description on $AdS_3\times S^3\times T^4$
factorises into three separate WZW models. The $AdS_3$ part gives rise to
an affine $SL(2,\R)\times SL(2,\R)$ left-right symmetric group manifold at
level $k>2$, which again determines the conformal charge of the $S^3$
theory on the group manifold $SU(2)$. In this section we chose $X$ to be $T^4$, 
because of its simple representation as an Abelian $U(1)^4$ model. Later on, 
the case of $K3$ will be more appropriate because of its simple cohomological 
structure, where the additional $\Z_2$ orbifolding has no effect on our reasoning.

Near the horizon of $AdS_3$ there are two alternative descriptions. The
RNS formulation has its advantage in the calculation of the particle spectrum,
whereas the quantisation of the Liouville Lagrangian \liv\ in the Green-Schwarz
representation gives a better understanding of the sigma model on $T^4$. But
in both cases the algebra ends up to describe a $(4,4)$ supersymmetric CFT. In the
WZW description, the fermions and the bosonic part of the currents of $AdS_3$
are denoted by $(\psi^A,k^A)$, while those corresponding to $SU(2)$ are
$(\chi^a, j^a)$. Here we adapted the notation from \commads\ to those of 
\seibergwitten\ for the sake of clarity, although the OPEs are not
completely identical. Now, the construction of the supersymmetric conformal
algebra is straightforward. From the complete algebra \commads, the
parts we will need are only the contributions of the energy-momentum tensor and the
supercurrent for $AdS_3\times S^3$. In the RNS formulation, the SCFT of $T^4$
is a free field contribution of conformal charge $c = 6$ and denoted
$T_{T^4}$ respectively $G_{T^4}$. With these simplifications, the $\NN=4$
algebra reduces to 
\eqn\alga{\eqalign{
&T_X={1\over Q_5}\left(k^A k_A-\psi^A \d\psi_A \right)+
      {1\over Q_5}\left(j^a j_a-\chi^a \d\chi_a \right)+T_{T^4} \cr
&G_X={2\over Q_5}\left(\psi^A k_A-
      {i\over 3Q_5}\epsilon_{ABC}\psi^A\psi^B\psi^C\right)+
      {2\over Q_5}\left(\chi^a j_a-
      {i\over 3Q_5}\epsilon_{abc}\chi^a\chi^b\chi^c\right)+G_{T^4}\;.}}
The analogous formulation in the Green-Schwarz description follows from
\final, where the four spinor fields on $AdS_3\times S^3$ are denoted by
$S^\mu$ in addition to the Liouville field $\phi$. 
\eqn\algb{\eqalign{
&T_\phi=-{1\over 2}\d S^\mu S_\mu - {1\over Q_5}j^a j_a -
      {1\over 2}\d\phi\d\phi +
      {1\over\sqrt 2}\left(
      \sqrt Q_5 - {1\over\sqrt Q_5}\right)\d^2\phi+T_{T^4} \cr
&G_\phi^\mu ={1\over\sqrt 2}\d\phi S^\mu -
      {2\over\sqrt Q_5}\eta^a_{\mu\nu}j_a S^\nu +
      {1\over 6\sqrt Q_5}\epsilon_{\mu\nu\sigma\rho}S^\nu S^\sigma S^\rho -
      \left(\sqrt Q_5 - {1\over \sqrt Q_5}\right)\d S^\mu+G_{T^4}\;.}}
In both cases the conformal charge is $c = 6Q_5$ and corresponds to the case
of $Q_5$ D5-branes with one D1-brane on its surface. This is of course not
very satisfying, because it fixes the possible number of D1-branes. Taking a
closer look at the Liouville part of \algb, the improvement term of the
energy-momentum tensor suggests a generalisation for $Q_1$ D1-branes in form
of a minimal conformal model 
\eqn\liv{{1\over 2}\d\phi\d\phi - Q\d^2\phi}
with Liouville charge 
\eqn\livc{Q={1\over \sqrt 2}\left(\sqrt\beta-\sqrt{1/\beta}\right)\;,}
for $\beta =Q_5/Q_1$. But, as we will
demonstrate in Section 3, even if such an extension exists, there would
still be not enough KK states to get a one-to-one mapping with the primary
modes and its descendants. To get a better understanding of this problem one
has to take a closer look at the higher twist modes for large $N=Q_5Q_1$.

\subsec{Missing States}

It was Vafa \vafapuzz\ who first pointed out a discrepancy between the
number of KK modes in the $AdS_3\times S^3$ supergravity background and the
number of chiral primaries and their descendants. An observation, which was
further analysed by de Boer \deBoera, who suggested a solution \deBoerb\ by
considering the ``exclusion principle'' as first observed in \exclusion. But,
although it is very reassuring to know that the particle spectrum for both
sides of the $AdS/CFT$ description coincide in principle, one has to ask to
what states the original missing states translate in the \algb\ description.
To answer this question, the first step is the understanding of the KK modes
in the context of representation theory. 

What makes the quantisation of $AdS_3\times S^3$ so easy, is the
description of both spaces as group manifolds. In Subsection 2.1 we
started by the formulation of the conformal field theory on the group
manifolds $SL(2,\R)$, $SU(2)$ and guaranteed a consistent supersymmetric
formulation by the vanishing conformal charge and a further GSO projection.
And, because the worldsheet of the two-dimensional CFT is a cylinder, the 
left-right modes decouple so that it was possible to simply ignore the left
moving part of the modes. Things are similar for the $AdS$
description, but with the supersymmetric analog of $SL(2,\R)$. The spherical
harmonics of $S^3$ are representations of $SO(4)/SO(3)$ or in the more
appropriate form $SU(2)\times SU(2)/SU(2)$, whereas the $AdS_3$ space
decomposes into the left-right symmetric group $SU(2|1,1)\times SU(2|1,1)$.
The representation of short and long multiplets transforming under this group
can be found in \guenadin\ and is reviewed in \deBoera. Now, the KK modes
have a complete description as short multiplet representations of
$SU(2|1,1)$, decomposed under the diagonal group $SU(2)$.

The short multiplets are of special importance as they contain the massless
spin 2 fields of the supergravity theory. And, because this is the highest
possible spin by KK compactification, the multiplication of these short
multiplets has a massless single particle state as highest spin state. 
For this reaosn, the complete set of chiral primaries of the CFT is obtained by
taking an arbitrary product of single particle states on the supergravity
side. This is exactly the case covered by \algb. But of more interest to us
is the multiparticle spectrum which caused the original puzzle of the missing
states \vafapuzz. These have been identified in \deBoera\ as non-chiral
primaries and are thus elements of long multiplets. The basic observation
was that the tensor product of two short multiplets does not contain a
new short multiplet only but various longer ones as well. In the next section, we
will analyse the structure and origin of the multiparticle spectrum from the
conformal field description and the supergravity point of view. 


\newsec{The Partition of Branes}

In this section we pursue the analysis of the multiparticle spectrum one
step further. A simple argument shows, that there is no generalisation of
\algb\ to D1-brane charges larger than one. But, nonetheless, a comparison
between the particle spectrum of the CFT and the supergravity theory reveals
the structure of the missing states and allows their general construction in
terms of Jack polynomials.

\subsec{Null States and Jack Polynomials}

In Section 2 we reviewed the effective action of the long string in an
$AdS_3\times S^3\times X$ background. The basic structure, which governs the
residual supersymmetric algebra, is a Liouville term of the form \liv\ with
improvement term \livc. For $Q_1=1$ this reduces to the low energy
description \algb\ with the coupling constants \seibergwitten
\eqn\coupling{\eqalign{
     g_C(\phi) =& \exp{\left({1\over\sqrt2}{-1\over\sqrt\beta}\phi\right)}\cr
     g_H(\phi) =& \exp{\left({1\over\sqrt2}
                  \left(\sqrt\beta-{1\over\sqrt\beta}\right)\phi\right)}
}}
for the short string of the Coulomb branch and respectively the large string of 
the Higgs branch. Because the coupling constants enter the vertex operators
as additional screening charges, they characterise the vacuum of the Liouville
theory. In the following, we will show that the generalisation to arbitrary
values of $\beta=Q_5/Q_1$ is wrong. To demonstrate the failure of this
description we take the mode expansion of \liv\ in the free field representation 
\eqn\livs{ \LL_n = {1\over 2}\sum_{n\in \Z}{a_{n+m}a_{-m}} - \alpha_0 (n+1)a_n}
with $\alpha_0 = {1\over \sqrt 2}(\sqrt\beta -\sqrt{1/\beta})$ and conformal
charge $c = 1 - 12 \alpha_0^2$. Instead of writing down the explicit from of
the screened vertex operators, it is sufficient to consider the highest
weight states of the Fock vacuum, represented by $|\alpha_{r,s}>$ with
\eqn\weight{\alpha_{r,s} = {1\over \sqrt 2}\left((r+1)\sqrt\beta -
                                               (s+1)\sqrt{1/\beta}\right)\;.}
The values of $r$ and $s$ parametrise the partition of D1-branes on the
stack of D5-branes restricted to 
\eqn\range{1\leq s\leq Q_5-1\hskip 2em {\rm and}\hskip 2em 1\leq r\leq Q_1-1}
for $\beta=Q_5/Q_1$. To compare these vacuum states with the results from
\coupling\ one has to take into account that the addition of the four spin
fields $S^\mu$ and the currents $j^a$ in the algebra \algb\ results in a
shift in $s$ and $r$ by two units. Therefore, it is useful to interpret the
additional screening charges for the string vertex operators \coupling\ in
the framework of the Liouville theory as the vacuum states at $s=0$ and
$s=Q_5$ for fixed $r=0$. But to keep things simple, we will stick to the
range \range\ for the values of $r$ and $s$ and keep in mind that the
interpretation as string couplings of \algb\ is related to the Liouville
vacuum state by a shift in the parameters. With this agreement, the Higgs
branch of the long string has the simple representation $|\alpha_{0,0}>$ and
the short string of the Coulomb branch corresponds to $|\alpha_{0,Q_5}>$.
Having the KK particle spectrum identified with the Liouville states of \liv, 
the origin of the missing states becomes more clear. 

The generalisation of the states $|\alpha_{0,s}>$ along \coupling\ is
obvious. But what about the other highest weight states $|\alpha_{r,s}>$,
defined by the relation $a_0|\alpha_{r,s}>= \alpha_{r,s}|\alpha_{r,s}>$?
The reducible vacuum states of the minimal model decompose under the null
vectors, of which the first four take the from 
\eqn\Astates{\eqalign{
|\chi_{1,1}>\;\sim &\;a_{-1}|\alpha_{1,1}>\cr
|\chi_{1,2}>\;\sim &{1\over 2}\left( a_{-2}+
                  2\sqrt{{\beta\over 2}}a^2_{-1}\right)|\alpha_{1,2}>\cr
|\chi_{2,1}>\;\sim &{1\over 2}\left( a_{-2}-
                  \sqrt{2\over\beta}a^2_{-1}\right)|\alpha_{2,1}>\cr
|\chi_{1,3}>\;\sim &{1\over 3}\left( a_{-3}+
                  3\sqrt{\beta\over 2}a_{-2}a_{-1}+
		  \beta a_{-1}^3\right)|\alpha_{1,3}>\cr
|\chi_{3,1}>\;\sim &{1\over 3}\left( a_{-3}-
                  {3\over 2}\sqrt{2\over\beta}a_{-2}a_{-1}+
		  {1\over\beta}a_{-1}^3\right)|\alpha_{3,1}>\;.\cr
}}
Here we omitted the normalisation factors which will be of no
importance in our discussion. The vacuum states $|\alpha_{r,s}>$ are
invariant under two symmetries, what reduce the infinite number of
possible states two a finite set. A shift in the parameters $r$ and $s$ of
the vacuum state by $|\alpha_{r-Q_1,s-Q_5}>$ has no effect and thus
justifies the reduction of the parameters to the range \range. So the
highest possible spin for $N = Q_5Q_1$ is exactly $N$ and at the same time the
origin of the stringy exclusion principle as found in \exclusion. The second
symmetry exchanges the parameter $\beta\to 1/\beta$ with an additional
sign change for all operator modes $a_n \to -a_n$. From a string
theory point of view, this inversion corresponds to T-duality along the four
compact dimensions of $X$. Up to now, we have assumed that the number of
D-branes $Q_5$ and $Q_1$ should be prime, so that the quotient $\beta$ would
be in one-to-one correspondence with their charges. But from now on we will
drop this condition and analyse, as a first step, the null states as functions
of $\beta$. 

The classical moduli space for $Q_5$ and $Q_1$ is simply the symmetric product
$X^{[N]}$. But the case that the two charges are not prime contradicts the
Liouville structure of \final\ and its representation
within the Green-Schwarz formulation. To be more specific, take the example
$Q_5=b\,Q_1$ or $\beta = b$, with the largest possible conformal charge $c=1$
obtained for $b=1$. At this value the moduli space degenerates and the
Liouville theory reduces to the free field representation of the Virasoro
algebra. The first null state, not defined for $\beta=1$ is $|\chi_{2,2}>$.
Another interesting value is $\beta =2$, for which the null states \Astates\
reduce to Schur polynomials. These functions enter the representation theory as the
polynomial ring of the symmetric group $S_N$ and thus generate a basis for
the moduli space $X^{[N]}$. To clarify the origin of the missing states in
this new framework, we will give a short introduction to these Schur
polynomials and its generalisation to Jack polynomials for arbitrary $\beta$.
Here and in the following, we will use the conventions of the review
\nakajimaa\ where further references can be found. 

A partition $\lambda=[\lambda_1, \lambda_2, \lambda_3, \ldots]$ is a
nondecreasing sequence of nonnegative integers for a finite number of
$\lambda_i\neq 0$. A different way of presentation is 
$\lambda=(1^{m_1}, 2^{m_2}, \ldots)$ with $m_k= \#\{ i| \lambda_i = k\} $. 
The two notations are distinguished by the
different type of brackets. Characteristic numbers of a partition are the
sum of integers $\lambda_i$, denoted by $|\lambda | = N$ and the number of
nonzero entries in $\lambda$, noted as the degree and the length
$l(\lambda)$ of a partition. It is often useful to give the vector 
$[\lambda_1, \lambda_2, \lambda_3, \ldots]$ a graphical interpretation as
Young diagrams, but we will not make use of them here. The ring of symmetric
functions with rational coefficions is denoted 
\eqn\Lam{\Lambda_N = \Q[x_1,\ldots ,x_N]/S_N\;,}
where the symmetric group $S_N$ acts by permutation on the variables. As
already mentioned in the introduction, it is always possible to embed a
partition in the infinite dimensional space $\Lambda_\infty$. The most
general representative of $\Lambda_\infty$ is the monomial symmetric
function or orbit sum
\eqn\monomial{m_\nu=\sum_{{\rm dist. perm.}}x_1^{\alpha_1}\cdots
                    x_N^{\alpha_N}\;,}
where the sum is over all distinct permutations
$\alpha=[\alpha_1,\alpha_2,\ldots]\leq \nu$ of entries in $\nu$ with
$l(\alpha)\leq N$. There are two distinguished partitions for each integer
$n$, the elementary symmetric function $e_n=m_{(1^n)}$ with $l(\alpha)=n$ and the
power sum $p_n=m_{(n)}$ with $l(\alpha)=1$. Because of the ring structure it
is possible to represent the functions $e_n$ and $m_\nu$ by the power sum
$p_n$, with the monomial symmetric functions recursively expressed in terms
of $p_n$ by
\eqn\recu{ p_i m_\nu = \sum_\mu a_{\nu\mu} m_\mu\;,}
where the summation runs over the partitions of $|\nu|+i$ and the coefficients
$a_{\nu\mu}$ counting the number of multiplicities of entries in $\mu$.
The elementary symmetric functions are expressed more easily by the
generating function 
\eqn\femprev{E(z) = \sum_{n=0}^\infty e_n z^n = \prod_{i=1}^\infty (1+x_iz)\;.}
At $z=1$ this has the structure of the tree-level amplitude of a single chiral
fermion \refs{\nakajimaa, \strong}. For completeness, we will give the
corresponding bosonic amplitude, too
\eqn\bosprev{H(z) = \sum_{n=0}^\infty h_n z^n =\prod_{i=1}^\infty{1\over(1-x_iz)}\;,}
with the complete symmetric functions $h_n$. They are related to $e_n$ by
$E(z)H(-z)=1$ and are complementary to each other. An alternative representation
in terms of $p_n$ is given by 
\eqn\reps{H(z)=\exp{\left(\sum_{n=1}^\infty{1\over n}p_n z^n\right)} 
                \hskip 1em {\rm and} \hskip 1em
E(z)=\exp{\left(\sum_{n=1}^\infty{(-1)^{n-1}\over n}p_n z^n\right)}\;.}
The generating functions have a structure similar to a scalar bosonic field 
\eqn\similar{\phi  = \sum_{-\infty}^{\infty} {1\over n}a_n z^n\;,}
with a zero mode $a_0$ still to be defined. 
We postpone this calculation to
Section 5, after a more detailed analysis of the moduli space. With a
redefinition of $z\to z/n$, the function $H(z)$  becomes the generating
function of the Schur polynomials of which the first representatives
are $P_0(x)=1$, $P_1(x)=x_1$, $P_2(x)=x_2+(1/2)x_1^2$ and
$P_3(x)=x_3+x_2x_1+(1/6)x_1^3$. Comparing these polynomials with the null
vectors $|\chi_{r,1}>$ shows, that the vacuum states \Astates\ correspond to
partitions of the type $\lambda=(r^s)$ only. But already for $N=3$ this
representation is not sufficient to represent all possible partitions.
From the example in \nakajimaa\ we know the three different orbit sums for
$N=3$
\eqn\summe{\eqalign{
m_{[1,1,1]} =& {1\over 3}a_{-3}-{1\over 2}a_{-2}a_{-1}+{1\over 6}a_{-1}^3\cr
m_{[2,1,0]} =& -a_{-3} +a_{-2}a_{-1}\cr
m_{[3,0,0]} =& a_{-3}\cr}}
of which only the first one has a counterpart in \Astates. The other two
partitions are Jack polynomials $J_\lambda(x;\beta)$. For the positive and
real number $\beta$ we define an inner product in the ring of symmetric
functions by
\eqn\normpower{<p_\lambda, p_\mu>=
               \beta^{l(\lambda)}z_\lambda\delta_{\lambda\mu}\;,}
where $z_\lambda = \prod k^{m_k}m_k!$ for the partition
$\lambda=(1^{m_1},2^{m_2},\ldots)$. With this normalisation, the Jack
polynomials are defined recursively by the Gram-Schmidt method
\eqn\jack{\eqalign{
 J_\lambda(x;\beta)=& \sum_{\mu\leq\lambda}u_{\lambda\mu}(\beta)m_\mu(x)\cr
u_{\lambda\lambda}(\beta) = c_\lambda(\beta) \hskip 1em {\rm and} & \hskip 1em
<J_\lambda(x;\beta), J_\mu(x;\beta)> = 0 \hskip 1em {\rm if} \hskip 1em
   \lambda \neq \mu \cr}}
with the normalisation factor 
\eqn\normjack{ c_\lambda(\beta) = \prod_{s\in\lambda}(\beta a(s)+l(s)+1)\;,}
where the vector $s\in\lambda$ is a point in the Young diagram with $a(s)$ arms
and $l(s)$ legs. The appearance of Jack polynomials is common for algebraic
integrable systems as eigenvalues of the Hamiltonian function. For the case
at hand, the system of null states resembles the Calogero-Moser model, as
already noted in \nakajimab, where this information has been used to compute
the Virasoro algebra for the homology of the Hilbert scheme. Before we pick
up this aspect in Section 4, let us mention one further characteristic of
the Jack polynomials, which will prove important for the construction of the
W algebras. 

Besides from the Hamiltonian, the Jack polynomials have a further symmetry,
generated by the Dunkl operators \vinet
\eqn\dunkl{D_i=\beta x_i{\d\over\d x_i}+\sum_{i\neq j}
          {x_i\over x_i-x_j}(1-K_{ij})\;,}
for $i=1,\ldots,l(\lambda)$ and the matrix $K_{ij}$ interchanging the
positions of $x_i$ and $x_j$. One interesting property of this operator has been
analysed in \vinet, where $D_i$ has been used to construct creation
operators $B_k^+$, whose action onto the trivial partition generate Jack
polynomials, as each operator adds one further column
of length $k$ to the Young diagram and thus has the property of shifting each
line by one step. In Section 5 we will construct similar creation operators to
generate the missing partitions of \summe\ from primary states. The
successive action of these shift operators takes us to the W algebras. 

\subsec{The Twisted States}
			
Having a hand on all possible partitions, we still have to show
that the missing states in the KK spectrum have a representation as elements
of the just introduced polynomials. For this we go back to an example given
in \deBoera\ and relate the non-chiral primaries with cohomology classes of
the moduli space. The possible products of the differential forms then have
a simple representation as Young diagrams. It is of no surprise that all
these states belong to the twisted sector and thus did not enter the
previous discussion. 

In the introduction we mentioned the relation of matrix theory and
compactification on the $AdS_3\times S^3$ background. The advantage of this
formulation in the low energy limit is its immediate interpretation as field
theory degrees of freedom in a six-dimensional spacetime \refs{\bfss, \dvv}. 
The description of a stack of $N=Q_5Q_1$ D-branes in the Higgs branch takes
the form of a field $X^\mu$ with $\mu=1,\ldots 6$, which can be written as a
diagonal matrix  
\eqn\matrix{ X^\mu  = Ux^\mu U^{-1}\;,\hskip 2em U\in U(N)}
after a $U(N)$ rotation. The eigenvalues of $x^\mu$ are then the
coordinates $x_i$ with $i=1, \ldots, N$. For an arbitrary partition
$\lambda=(1^{m_1}, 2^{m_2}, \ldots)$ the matrix becomes diagonal after a
$U(1)^{m_1}\times U(2)^{m_2}\times\ldots $ rotation with new eigenvalues
$\tilde{x}_i$ for $i=1,\ldots l(N)$. The comparison of this matrix description with
the Dunkl operator, identifies the coordinates of \dunkl\ with the
eigenvalues $\tilde{x}_i$. It is not difficult to understand the origin
of the noncommutative structure of matrix theory. To start with the simplest
possible example, decompose the hermitian matrix $X^\mu$ into Fourier modes
of the matrices $U,V\in U(N)$, satisfying the relations
\eqn\relation{ U^N=V^N=1 \hskip 1em {\rm and} \hskip 1em UV=qVU\;,}
with $q=\exp{(2\pi i/N)}$ \bfss. The two group elements are the generators of
$SU(2)$ embedded into $U(N)$, why the Fourier modes depend on
two variables, corresponding to the two-dimensional space the root vector
lives in. Now the matrix field $X$ can be written as
\eqn\fourier{X =\sum_{n,m}x_{nm}U^nV^m}
or alternatively in the lattice of the root space
\eqn\coeffi{x(p,q) = \sum_{n,m} x_{nm}e^{{2\pi i\over N}(np+mq)}\;,}
where the noncommuting coordinates $(p,q)$ determine the length of the
weight lattice. From this, the
generalisation to groups of higher rank and different type is obvious and
well understood. At least for simply laced algebras, the suggestive form
\coeffi\ has the generalisation 
\eqn\blabla{f(x) = \sum_{w\in W(R)} f_w q^{<w,x>}\;,}
with $W(R)$ the weight space of the roots $R$ embedded into $U(N)$. The
noncommutative structure of the matrix theory introduces the new
parameter $q=\exp{(2\pi i/N)}$ and generalises the Jack $J(x;\beta)$ to
Macdonald polynomials $J(x;t,q)$. But in the following we will only adopt
the formalism of commutative geometry and set the value of $q$ to one. To do so,
one first has to set $q=t^\beta$ and then take the limit $t\to 1$. The exact
relation thus follows from \vinet
\eqn\relat{J_\lambda(x;\beta)=\lim_{t\to 1}
          {J_\lambda(x;t,q=t^\beta)\over (1-t)^{|\lambda|}}\;.}

Apart from the noncommutative structure, matrix theory gives an efficient access 
to the explicit construction of the higher twist states in the Green-Schwarz formulation
\dvv. It is a special feature of two-dimensional supersymmetry that the
Lagrangian not only consists of single particle multiplets, but also contains
twisted multiplets, which play a key role in the calculation of the
Bekenstein-Hawking entropy \refs{\macentropy, \micentropy, \entropywit}. The
only drawback of the Green-Schwarz formulation is the discrete light cone
limit, analogous to the large $N$ limit of the $AdS/CFT$ description. But,
as stated above, the Liouville action not only gives a valid description of
the large string but also of the short string \seibergwitten, and there is
no reason to assume that things will be different in the twisted sector. The
primary twisted states are formulated in terms of the $\NN=(2,2)$ topological
conformal field theory and therefore have a simple interpretation as
cohomology classes of the orbifold $X^{[N]}$. Here we will not go into the
details, but refer to \refs{\exclusion, \commads} for an explicit construction 
and further analysis. For the special case $N=1$ the orbifold simply reduces
to the original space $X$, but now with the cohomology classes of $X$ as
primary states. The translation of a $(p,q)$ cohomology form $\omega^A$
and its corresponding primary state in the $(a,c)$ ring takes the form
\exclusion 
\eqn\cohom{ \Phi^A = \omega^A_{ab\ldots\bar a \bar b \ldots}
                    \psi^a(z)\psi^a(z)\ldots
		    \psi^{\bar a }(\bar z)\psi^{\bar a }(\bar z)\ldots\;.}
As has been observed by Vafa and Witten in \strong, the inherited
commutation relations of the even or odd numbers of spinors have a simple
representation as spinor or bosonic creation operators $\alpha^A_{-1}$, with
the familiar commutation relations 
\eqn\comm{\{\psi^A_{n},\psi^B_{m}\}=\delta^{AB}\delta_{n+m} \hskip 1em
          {\rm and} \hskip 1em
	  [\alpha^A_{n},\alpha^B_{m}]=n\delta^{AB}\delta_{n+m}\;,}
after a convenient rescaling of the operators by $1/\sqrt{Q_5Q_1}$. Here the
index of the ``spacetime'' runs up to $\# H^*(X,\Z)$. It is interesting to note
that an analogous relation for the untwisted sector has been found in
reference \commads, with the spacetime index parametrising the four compact
dimensions of $X$. Assuming that both algebras are
independent from each other, similarly to the two-dimensional field
description. But this seems to be somewhat unnatural as
the conformal field theories have the conformal charges $c=6Q_5Q_1$
and $c=6Q_5$. So that for $Q_1=1$ both descriptions are
allowed, which is in contrast to the result found above, that the large
string is part of the untwisted sector and not of the twisted one. 
A possible solution to this problem is the introduction of a
double index for the operators $\alpha^A_{N,Q_1}$, which takes care of the
operator mode $N=Q_5Q_1$ as well as of the number of D1-branes $Q_1$ with
corresponding conformal charge 
\eqn\confcharge{c=c(N,Q_1,\# H^*(X,\Z))\;,}
which has the property to change by a factor of 6 for $K3$ (and 4 for $T^4$),
if the D1-brane charge exceeds the boundary of $Q_1=1$.

Up to now, we have only focused on the primary states of the twisted
sector, which are the cohomology elements of $X$ and its tensor products.
The introduction of further D1-branes changes this picture as additional
singular regions of the orbifold are blown-up and increase the dimension of
the moduli space. The blow up modes of the singular CFT are marginal
deformations of the twisted sector, whose number for the $D5/D1$ brane
system are easily calculated, because all states of higher twists than $\Z_2$ are
irrelevant. This results in a drastic reduction of the possible Young
diagrams, as for a stack of $N=Q_5Q_1$ branes only the two partitions
$\lambda_1=(1^N)$ and $\lambda_2=(1^{N-2}, 2^1)$ are related to marginal
deformations. Instead of an explicit construction of the vertex operators
\refs{\commads, \dvv}, we will use the more efficient description of the
Liouville theory in combination with the shift operator \dunkl\ in
Section 5. Following the above example \deBoera, there are 24 cohomology
elements of the primary states and two further ones from the orbifold
$K3^{[2]}$. Knowing that all non-chiral primaries are tensor products of the
differential forms, they can be identified with the two partitions $\lambda_1$
and $\lambda_2$ and its reduced tensor products. Introducing an additional
number $Q_3$ of  D3-branes along the lines of 2.2 reduces the conformal 
weights of the twist operators by $1/ Q_3$ and a thus larger number of
marginal deformations have to be taken into account. The
consideration of D3-branes is only one example, where the explicit construction
of the vertex operators in terms of the CFT shows up the limits of an
analysis along these lines. 


\newsec{The Moduli Space}

In the previous paragraph we made use of the Virasoro algebra of the free field
realisation of the Liouville theory for the twisted and untwisted sector.
From a mathematical point of view this construction has been known for a
long time \refs{\nakajimaa, \nakajimab, \nakajimac} and entered the physical
discussion in \strong. In connection with the McKay correspondence it
successfully explained, why affine Lie algebras arise after quantising
gauge theories. But this is of minor interest here. In this
section we will show how the previous $AdS_3/CFT_2$ discussion can be
related to the more general case of intersecting branes in a flat background
space. The first part of this section therefore analyses the moduli space of
intersecting branes. A special example thereof is the matrix interpretation of M
theory with the DBI action as the string theoretic approximation. But the 
consequences of the quantisation are not considered until the second part. This
leads to the Virasoro description of the underlying moduli space and
introduces the anomaly term of D-branes. The fusion of the DBI action with
the Chern-Simons term as the D-brane charge anomaly will be the final result
of Section 5. 

\subsec{The Coulomb and Higgs Branch in Field Theory}

The D-brane action contains basically two superficially different parts, one
is the DBI action or kinetic energy of the brane, while the second
contribution cancels the anomalous D-brane charge by a Chern-Simons term.
As already noted in the introduction, quantising the DBI action is an
unsolved problem, but even worse, the expansion of the determinant makes a
perturbative analysis impossible. A more appropriate approach was
found in \bfss\ using the language of matrix theory. In the last section we
already introduced the 1+1 dimensional $SU(Q_5)$ gauge theory with $Q_1$
instantons and the $AdS_3/CFT_2$ correspondence as alternative effective
$D5/D1$ brane description. Whereas the matrix description proved to give an
elegant connection between representations of the symmetric group $S_N$ and
the twisted states. For this discussion one must not forget that each
representation corresponds to one specific D-brane partition. This way, matrix
theory allows a simple mapping between the geometric picture of
intersecting flat branes and an effective gauge theory with supermultiplets.
Although the quantisation of both formulations is not known, the field
theoretic description allows a simple investigation of the underlying moduli
space \refs{\smallinst , \modulid, \hananywitten, \solwitten} and gives a better
understanding of the creation operator $B^+_k$. Our main interest in this
article is the $D5/D1$ brane system, which fixes our discussion to the type
IIB sector in string theory, but is no principal limitation. In terms of
\bfss, the potential of the matrix action is described by the
Lagrangian 
\eqn\pot{ L = {1\over 2\pi\alpha'^2 g}\tr\left( {1\over 4}[X^\mu, X^\nu]^2 +
              {1\over 2}\psibar\gamma_\mu [X^\mu, \psi]\right)\;,}
where the noncommuting matrix fields $X^\mu$ are representations of $U(N)$
as in the previous section. The Fourier decomposition \fourier\
introduces an infinite number of modes $x_{nm}$ and consequently infinitely
many terms contributing to the kinetic energy of the brane. To recover the
classical DBI action, one only has to expand the commutator relations of the
matrix fields to first order. The case of interest is the D1-brane
for which the expansion of $N$ separated single branes reduces to
\eqn\commrela{\eqalign{
[X_0, X_1] = 2\pi i (1+2\pi \alpha'F_{01}+\ldots ) \;, \hskip 2em &
[X_\mu, X_n] = 2\pi i D_\mu X_n+\ldots   \cr
[X_\mu, \psi] = 2\pi i D_\mu\psi+\ldots \;, }}
with the index $\mu$ running over the bulk coordinates and the spacetime
index $n$ of the D1-brane. Further powers of $\alpha'F_{01}$ are contained in
the higher order terms of the expansion. The leading term of
\pot\ is proportional to the surface density of the brane and reproduces the
DBI action
\eqn\dbi{ S = T_p\int_{\d X}\Tr\sqrt{\det\left[G_{nm}+2\pi\alpha'F_{nm}\right]}+
               \ldots \;,}
where we have set the NS B-field to zero. Next, we separate the metric term
$G$ from the determinant and rename the residual curvature $\FF=G^{-1}F$.
The Lagrangian then takes the simple form 
\eqn\class{\Tr\sqrt{1+\FF}}
under the additional assumption $2\pi\alpha'=1$. Written in this simple form
the DBI action reveals its topological structure, depending on three
contributions. 

The derivation of \class\ started from the assumption of $N$ separated
branes with a $U(1)$ gauge freedom on each surface. We are not able to
show from matrix theory alone that an analogous formula holds in the case
of higher gauge groups, but we will give an indirect argument at the
end if this subsection that in principle the structure is still correct. It is
possible to generalise $\FF$ formally to higher gauge groups. The
first thing we learn from \class\ is that the discriminant is the
Chern class $c(\FF)$ of a gauge bundle with structure group ranging from $U(1)$
to $U(N)$. Consequently, the square root expresses the Chern class of a
single D-brane. Probing the stack of $N$ separated branes by the interaction
with an external brane follows now from the relation
$\chern(\FF+\GG)=\chern(\FF)\chern(\GG)$. A special case is $\FF = \GG$
where the square root of \class\ reduces to the simple Chern class
$\chern(\FF)$. In principle, this form also allows interactions between
branes of the stack itself, which leads us to the trace over the Chan-Paton
factors. Branes interacting among themselves increases the rank of the gauge
group and thus the rank of the underlying vector bundle, but it also reduces
the number of possible combinations of the probing brane with the stack.
Taking into account that we have set the B-field to zero and 
the branes are free to move relatively to each other. So the action
\class\ is only the first term in a series of different partitions, where
the first part corresponds to the Coulomb branch $\lambda = (1^N)$ of the
last section with one D1-brane interacting with a stack of $N$ D5-branes.
Now, the complete action along the lines of \class\ corresponds to the Chern
class of all partitions of branes, which can formally be written as
\eqn\formally{
S=\int_{\d X^{[N]}} \sqrt{\chern(u)}\hskip 1em {\rm with} \hskip 1em
            u \in K(X,\Z)}
and $\chern(u)\in\End(H^*(X^{[N]}))$ as Chern class of the Hilbert scheme at
order $N$. As the D-branes are wrapped around the Hilbert scheme $X^{[N]}$,
the integral has to be evaluated over its boundary $\d X^{[N]}$. In the next
section we will demonstrate, how the integral over the ``boundary of the Hilbert
scheme''  is to be calculated recursively in the framework of the Virasoro
algebra. Unfortunately, we have no representation in terms of gauge fields
similar to the D-brane anomaly, but knowing the exact kinetic energy is
always the first step of perturbation theory, we will use the propagator
$c^{-1}(u)$ to calculate the residual brane dynamics in Section 5. But
neither the matrix description nor the $AdS/CFT$ correspondence alone provides 
an applicable method to do this.

The DBI action restricts the analysis of the moduli space to the Coulomb
sector and leaves the more general Higgs branch and the intermediate states
completely untouched. Furthermore we still have to show that the assumption of
\formally\ is correct. But it is not very reasonable to try and generalise the
DBI action. Instead we will combine the ADHM description of intersecting
D-branes in matrix theory \modulid\ with the Virasoro algebra
\refs{\nakajimaa, \nakajimac}.
This way one avoids the expansion of matrix fields in terms of gauge
interactions \commrela\ and thus the problem of analysing the DBI action,
which is correct only in case of the Coulomb branch of the underlying field
theory. Basically the interpretation of intersecting matrices and their
moduli spaces has been done in \refs{\modulid, \neks} for the commutative as
well as the noncommutative case. Therefore we will not repeat the actual
correspondence and instead start with the relevant mathematical formulation
of moduli spaces as hyperk\"{a}hler moment maps. As a four-dimensional
example consider the space $\IC^2$ with the Hilbert scheme $(\IC^2)^{[N]}$
as the moduli space of zero-dimensional subschemes \nakajimab. Because $K3$
has a local representation as an ALE space, this example can be seen as an
approximation to the above problem. Following the ADHM construction of the
moduli space of charge $Q_1$ instantons with $SU(Q_5)$ symmetry, one
introduces two matrices $B_1, B_2\in \Hom(V,V)$ and two vectors
$I\in\Hom(W,V)$ and $J\in\Hom(V,W)$ in the complex hermitian vector spaces
$W=\IC^{Q_5}, V=\IC^{Q_1}$. Now, the actual moduli space  $\MM(Q_5,Q_1)$ is
determined by the set $(B_1, B_2, I, J)$
\eqn\mathrelm{\eqalign{
\mu_{\R} & = {i\over\sqrt 2}\left([B_1,B_1^+]+[B_2,B_2^+]+II^+ + J^+J\right)\cr
\mu_{{\bf C}} & = [B_1, B_2] + IJ\;,}}
where the space $\MM$ is defined as the $U(V,\IC)$ invariant space of 
\eqn\mathmod{\MM(Q_5,Q_1)=
             \left(\mu_{\R}^{-1}(0)\cap\mu_{{\bf C}}^{-1}(0)\right)/U(V)\;}
with the group $U(V)$ acting on $(B_1, B_2, I, J)$ by the relation
$g\cdot(B_1, B_2, I, J)=(gB_1g^{-1}, gB_2g^{-1}, gI, Jg^{-1})$. For a
further discussion of these equations we refer to
\refs{\nakajimab, \modulid, \neks} and references therein. The analysis of
the moduli space with respect to the vector $J$ brings us to the
consideration of noncommutative geometry, which is not of primary interest
in the discussion here. We will set $J=0$ for the time being and
postpone the discussion to a later section. What is important at the moment
is the connection between the matrix description of M theory and the moduli
space determined by the ADHM description. As we are not interested in
special solutions of \mathrelm, it is sufficient to classify the possible
solutions without actually calculating their matrices \nakajimaa.
In a first step we decompose the holomorphic vector space $V$ into its
weight spaces by the torus action $\lambda:T^2\to U(V)$, satisfying the
conditions 
\eqn\torus{\eqalign{
t_1B_1 & = \lambda(t)^{-1}B_1\lambda(t)\cr
t_2B_2 & = \lambda(t)^{-1}B_2\lambda(t)\cr
I & = \lambda(t)^{-1} I\;,}}
with the two coordinates $(t_1,t_2)\in T^2$ of the torus. By decomposing the
matrices into $B_1, B_2$, the vector space $V$ separates into
the weight spaces $V = \sum_{k,l} V(k,l)$, defined by 
\eqn\weights{V(k,l)=\{v\in V|\lambda(t)\cdot v = t_1^kt_2^l v\}\;.}
In this weight space the defining relations \mathrelm\ of the moduli space
take the simple form
\eqn\simplerelm{\eqalign{
B_1: & \; V(k,l) \;\to\; V(k-1, l)\cr
B_2: & \; V(k,l) \;\to\; V(k, l-1)\cr
 I : & \hskip 2em W      \;\to\; V(0,0)\;.}}
Since we set $J=0$, the matrices satisfy the commutation relation
$[B_1,B_2]=0$ as a consequence of the second equation of \mathrelm. In
combination with the finiteness of the vector space $V$, the operation of
the matrices cuts a finite grid out of the two-dimensional net of the weight
spaces $V(k,l)$, which in \refs{\nakajimaa,\nakajimac} has been connected to Young
diagrams, the graphical representation of partitions
$\lambda=[\lambda_1,\lambda_2,\ldots]$ as introduced in Section 3. Actually,
the decomposition of $V$ into weight spaces is the abstract formulation
of the matrix expansion \commrela\ with the square root of the Chern class
of $V$ as the generalisation of the DBI action. The Coulomb branch for
example decomposes into $0\to V(0,0)\to V(0,1)\to\ldots\to V(0,N)\to 0$
with each space of dimension one and Chern class $\det(1+\FF)$. 

The discussion of matrix theory and its moduli space joins the results
of the $AdS_3/CFT_2$ discussion of the last section, where we have shown
that the null vectors of the Liouville theory describes on the one hand the
partitions of branes and on the other hand the cohomology ring of the
Hilbert scheme in terms of a Virasoro algebra. Therefore we argue, that the
combination of both theories allows the description of M theory not only in
the large $N$ limit or at the boundary of the $AdS$ space but for all values
of $N$. Here we have to mention, that the discussion of K-theory for D-brane
/ anti D-brane interactions suggests a small limitation to this assumption,
as the statement is only correct for $N\geq 10$. We believe that this
restriction has the same origin as the $Q_1$ independence of the conformal
charge, a puzzle we cannot solve so far. 

\subsec{Hilbert Schemes and the Virasoro Algebra}

In the previous section we have shown that the moduli space of the multiplets as
determined by matrix theory are intimately connected with the Virasoro
algebra representation of Young diagrams. In this section we will tie some
of the loose ends together, as the connection between the degree of a
partition and the anomaly condition of D-branes or the signature of
spacetime for the Virasoro algebra and the moduli space. The results
entering our discussion are based on \refs{\nakajimab, \ktheoryc} and we
will repeat some of the ideas in the context of our analysis of D-brane
interactions. 

For our generalisation of the Green-Schwarz description of the $D5/D1$ stack
we claimed that the spacetime dimension of the Virasoro algebra of the
twisted sector is determined by the number of cohomology elements of the
manifold wrapped by the D5-brane \cohom. But there is a slight difference
compared to the mathematical point of view, where the dimension is fixed by the
Neron-Sev\'{e}ri lattice. As an explicit example consider the case $X=K3$.
The dimension of this lattice is basically the number of $H^{1,1}(X, \Z)$
elements, which is $20$ for $X=K3$, 
but the whole number of cohomology elements is $24$.
Surprisingly, string theory combines the information about the
Neron-Sev\'{e}ri group with the number of D-branes and the gauge group of
the string theory into the D-brane anomaly \ktheoryc. For the example of
$K3$ this can be seen as follows. The intersection form of $H^2(K3, \Z)$ is
isometric to $((-E^8)\oplus H)\times((-E^8)\oplus H)$ and defines the gauge 
group of the effective (bosonic) string theory in $24+2$
dimensions. As long as the considered manifold has a complex structure, the
Hodge decomposition of the $H^{2}(\Z)$ forms is related to the decomposition
under the complex structure, since the antiselfdual cohomology elements are
related to $H^{1,1}(\Z)$. At least for a stable moduli space, the lattice of
selfdual RR fields \ktheoryb\ is now sufficient to determine the
Neron-Sev\'{e}ri lattice. But the WZW term of the D-brane anomaly introduces
the number of intersecting D-branes as further information. Although the
example of $K3$ is interesting, it is not very convincing for our problem of
type IIB string theory, as we claimed that the twisted sector and the
untwisted one belong to the same algebraic construction and this forces a
ten-dimensional space $(1,9)$ with conformal charge $12=8+4$ and not the
$24+2$ dimensions of $K3$. We already mentioned the problem with the
D1-brane charge before. This puzzle can be reformulated in terms of the
moduli space of $X$ and we believe to have found one possible solution by
Enriques surfaces $Y$. The universal covering of $Y$ is $K3$ with
intersection form $(-E^8)\oplus H$. Even the cohomology groups 
$H^2(Y,\Z)=\Z^{10}\oplus \Z_2$, $h^0 = h^4=1$
determine the underlying string theory to be ten-dimensional.
Although it is very speculative, we 
think that the number of D1-branes determines the number of $Y$
coverings. As the fundamental group of $Y$ is $\Z_2$, this reproduces the 
ten-dimensional string theory for $Q_1=1$ and at $Q_1=2$ the manifold is
basically $K3$. A further hint comes from the W algebra analysis itself, as
the canonical bundle of $Y$ is of torsion class $K_Y^2=\OO_Y$. It is
important to note that D-branes cannot wrap Enriques surfaces, but we think of 
it as a generating space for the effective string theories. 

Before we turn to the discussion of the Virasoro algebra, some further
comments on the Neron-Sev\'{e}ri lattice $\NS(X)$ are in order as they
uncover some new aspects of the D-brane anomaly and the quantisation of
selfdual RR fields. As assumed before, $X$ is a complex manifold with
structural sheaf $\OO_X$. It follows from the exponential map
$\exp:\OO_X\to\OO_X^*$ that the sequence 
\eqn\expseq{0\to\Z\to\OO_X\to\OO^*_X\to 0\;,}
gives rise to the cohomology sequence 
\eqn\cohoexpseq{\to H^1(X,\Z)\to H^1(X,\OO_X)\to H^1(X,\OO_X^*)\to
                    H^2(X,\Z)\to H^2(X,\OO_X)\to\;,}
where the derivative $d:H^1(X,\OO_X^*)\to H^2(X,\Z)$ defines the Picard
group of $X$. If we define $\Pic_0(X)=\Ker(d)$, the N\'{e}ron-Severin group is
isomorphic to the quotient 
\eqn\nsg{\eqalign{
\NS(X) &  =  \Pic(X) / \Pic_0(X)\cr
       &  =  \NS(X) / \Tors(\NS(X))\times \Tors(\NS(X))\cr
       &  =  \NN\SS(X) \times \Tors(\NS(X))\;,}}
where we have separated the torsion part of $\NS(X)$. For the following it is
important to get a better understanding of the significance of the
N\'{e}ron-Severin group. Therefore we will first give a simple example. An
appropriate starting point is the geometric engineering of string theories.
Consider for example the ALE space $X=\IC^2/\Gamma$ with
nondegenerate lattice $\Gamma=v_1\Z+v_2\Z+v_3\Z+v_4\Z$ cutting out the curve
$C=\sum a^{ij}S_{ij}$ generated by the cycles $S_{ij}$. Choosing a
differential form $\omega$ on $X$, we get for each of the four cycles 
\eqn\calcone{ \int_{S_{ij}}\eta = \alpha_i\beta_j-\alpha_j\beta_i\;.}
Thus for the curve $C\subset X$ we get
\eqn\calctwo{\int_{C}\eta
             =\sum_{i<j}a^{ij}(\alpha_i\beta_j-\alpha_j\beta_i)=0\;.}
The solutions of this relation determines the Picard lattice. Here the
answer is quite trivial as $\Gamma$ is nondegenerate and thus
$\Pic(X)=\Pic_0(X)$. But in general there are nontrivial solutions
generating the $\NS(X)$ lattice. The skew symmetric product entering the
above calculation is known in field theory as the Dirac quantisation
condition and relates two elements in the homology. The subgroup $\Pic_0(X)$
determines exactly those elements for which no duality relation
exists. But if the N\'{e}ron-Severin group is non trivial some of the numbers
$(\alpha_i, \beta_i)$ have to vanish. The result \calctwo\ takes a rather
suggestive form if one identifies the numbers $(\alpha_i, \beta_i)$ with its
cohomology elements in $H^{2}(X,\Z)$ and the differential form $\eta$ with
the $\theta$-term in gauge theory $F*F$. In string theory this intersection
form is known as the anomaly condition
\eqn\inter{ <x,\bar y> = \int_X G(x)\wedge *G(\bar y)
                            = \int_X G(x)\wedge G(y)\;,}
but as shown in \ktheoryc\ K-theory always combines the anomaly term with
the kinetic energy of the RR fields
\eqn\teta{ \Theta(x,y) = \exp{\{(x,y)\}} = \exp{\{<x,y> + \tau <x,\bar y>\}}\;,}
into a theta function $\Theta(x,y)$ for $x,y\in K(X)$, where we
understand the gravitational anomaly as included. This theta function
describes more than just the Picard lattice of $X$. Actually, it combines
the degree of D-brane partitions, of which we studied the example $N=Q_5Q_1$
with a generalisation of the NS lattice, as the gravitational anomaly
introduces a further $\Z_2$ dependence. The knowledge of the theta
function \teta\ already determines the topological information we will need
for the construction in the next section. But before we turn to the
algebraic aspects of the moduli space, let us emphasize at least two
important aspects of \teta.

The main motivation for the introduction of the NS lattice $\Gamma$ was its
appearance as the metric of the spacetime in the Virasoro algebra. Now the
action of S-duality maps the lattice to its inverse $\Gamma^*$ and
T-duality, as it interchanges the RR fields in \teta, maps it to its
transpose $\Gamma^\Tr$. Furthermore, the intersection \inter\ at level zero,
$N=0$, determines the conformal charge of the Virasoro algebra and since the
$AdS/CFT$ correspondence is correct only in the large $N$ limit, the only
information we get from the field theoretic description of M theory is that
of an $N$-fold covering of the manifold $X$ but not of $X$ itself. This again
underlines the necessity to understand M theory for finite $N$, because any
manifold $X$, whose infinite covering is isomorphic to $K3$, gives an
equivalent description in the field theory formulation. The class of
Enriques surfaces we mentioned above is one example.

The spacetime dimension of the Virasoro algebra is actually only the first
step in the analysis of the moduli space. To extract information from this
construction one still has to find a way as to combine the modes in such a
way to relate the topological information to the algebra. This is simple
enough from the point of string theory, as the Virasoro algebra is only an
intermediate step in the process of quantisation. The elliptic genus and the
partition function of the torus $T^2$ are only two examples of how to extract the
topological information from the algebra. The by far more general objects
are the vertex operators of fields, which we now turn to. Here again, it is
interesting to note how similar the mathematical description is when compared to
the physical one. The Virasoro modes of the physical fields have to respect
additional conditions originating from compactifications or boundaries,
whereas the moduli space determines a kind of effective Virasoro algebra, which
again defines a ``field''. How to calculate this effective algebra is the
subject of the next section. Here we will only give an idea of the
mathematical construction behind the vertex operator. Its formal definition
is simply the infinite sum \nakajimab
\eqn\mathvertex{\eqalign{
V_{\NS(X)} & = \sum_{c_1\in\NS(X), ch_2} H_{\NS(X)}(\MM(c_1, ch_2))\cr
           & = \sum_{n}H_{\NS(X)}(X^{[n]})\otimes \Q[\NS(X)]\;,}}
where the order of the Hilbert scheme $X^{[n]}$ is determined by 
\eqn\orderhilbert{n=-{\rm ch}_2(X)+{1\over 2}<\chern_1, \chern_1>_X\;.}
As shown in \harvmoore\ this is the product of the Mukai vector with the
canonical sheaf $\OO_X$ or simply the bundle of wrapped D5-branes,
intersecting with the line bundle of D1-branes. Written in terms of the
D-brane anomaly \inter\ it takes the form
\eqn\dfsfs{N = {\chi(X)\over 24}-{1\over 2}\int_X G\wedge G\;,}
which is related to \orderhilbert\ by $n=-24\; N$. One further interpretation
of the anomaly number $N$ originates from the K-theoretic picture. An alternative
way to write the D-brane anomaly \inter\ is by insertion of the equation
of motion for the RR fields with the result
\eqn\interalt{<x,\bar y> = \int_X \hat{A}(X)\ch(x)\ch(y)\;,}
where the elements $x\in K(X)$ are represented by vector bundles. Take for
simplicity $x\in [E]$, the class of D5-brane bundles and $y\in L$, the class of
D1-brane line bundles. The degree of the D-brane partition $N$ is then
determined by the monodromy around the effective divisor $D$ of the line
bundle on $X$ defined by a shift $[E]\to [E]\otimes\OO_X(D)$. As long as
the D-branes satisfy the BPS bond, both formulas \inter\ and \interalt\ give
an integer anomaly number $N$. As is well known, this is no longer true if a
selfdual NS B-field is turned on, corresponding to a shift in the field tensor
$F\to F + B$. Moving around the divisor
takes the anomaly number $N$ into non integer values and thus contradicts our
assumption that $N$ is the degree of a partition. But we believe that this
problem has a simple solution. As for $B^+\neq 0$, the system of interacting
D-branes are no longer BPS, the winding numbers around $X$ and the singularity of the
D1-brane $D$ need not to be integer any more. From \inter\ we know that the
RR fields generate a two-dimensional torus similar to the Picard torus. Let
$[C_5]$ and $[C_1]$ be the two cycles of this torus with a foliation
$[C]=Q_5[C_5]+Q_1[C_1]$ in the case of BPS branes. Turning on a B-field,
the cycles do not close, up to the noncommutativity parameter $\theta$.
For the same charges the homology element $[C]$ now becomes
$[C]=Q_5[C_5]+Q_1[C_1]+\theta Q_5Q_1 [C_5\cap C_1]$. It is therefore necessary to
move back along the path by an amount of $\theta N [C_5\cap C_1]$. But moving
in the opposite direction is equivalent to the monodromy around the inverse
line bundle $L^{-1}(\theta)$ represented by an anti D1-brane of charge $\theta$. In the next
section we will make use of this construction. But before that, we have to explain
how moduli parameters such as $\theta$ are to be incorporated into the algebra.

This brings us back to the effective Virasoro algebra of the moduli space.
One example already entered the discussion in the analysis of the partition
of the $D5/D1$ stack \similar. After the by now common sign change in the
Virasoro modes, the commutation relations take the form \nakajimaa
\eqn\topcomm{ [q_n^i, q_m^j]= (-1)^n n\delta^{ij}\delta_{n+m}\;.}
The calculation of this effective algebra was simple, because of the
analogy between the allowed partitions of the stack of D-branes and the
generating function of the elementary symmetric functions \reps. As we will
see in the next section the similarity between \femprev\ and the Chern
classes of the moduli space is no coincidence. Actually, the connections
between string theory and the algebra of vertex operators are far more
numerous and we refer to \nakajimab\ and references therein.


\newsec{W Algebras and Hilbert Schemes}

In this chapter we will finally construct the shift operator analogous to 
\dunkl\ in creation modes of the free field representation of the Liouville theory.
Following the discussion of Section 3, the bosonic part contains the basic
structure, so that as a first attempt we can restrict the discussion to the case of 
$K3$. We will make two important assumptions here, first the noncommutative part of
\dunkl\ will be omitted and furthermore the spacial dimensions of the
operation modes and thus of the operators itself are not indicated in the
formulas. The last assumption is made because the correct structure of the
conformal charge is not known to us for general D1-brane charges and thus
has to be omitted. But the formulas will be true for the untwisted as well
as the twisted sector separately so that the results make sense in both cases. 
With this understood we begin with the analysis of the shift operator and its
generated W algebra. It is then a simple task to generalise this result to
the more general supersymmetric case of a $W_{\infty}(\lambda)$ algebra,
which finally includes the complete cohomological structure of the twisted
sector even for manifolds different from $K3$. As a test that all states of
Section 3 are reproduced correctly, we will reconsider the simple case of a
stack of $D5/D1$ branes and calculate the vertex function and the
corresponding commutation relation of the resulting operator modes \topcomm.
As an outlook of the advantage of this formulation we consider the three point
amplitude for one D3-brane inserted into the stack of $D5/D1$ branes, a
result which has been calculated by M. Lehn \lehn. Finally we
speculate on a correspondence between six-dimensional Calabi-Yau manifolds
and $W_3$ algebras.

\subsec{The $W_{\infty}(\lambda)$ Algebra}

In Section 3 we introduced Jack polynomials $J_\lambda(x;\beta)$ to compare
the null vectors of minimal models and its descendants with representations
of the symmetric group $S_N$. These specific functions are the solutions of
the Calogero-Moser model and have the Dunkl operator \dunkl\ as a kind of covariant
derivative. As the analysis of \vinet\ shows, Jack polynomials can be
constructed by creation operators $B^+_k$ from the Dunkl operator as
symmetry generating operator that defines the corresponding Hilbert space.
How are the Hilbert spaces of the minimal model and the integrable
system related, and what is the connection between the coordinates $x_i$, the
string modes $a_n$ of \livs\ and the eigenvalues of the matrices \relation\
in matrix theory?

In a first step, to identify the relation between the coordinates with the
matrices \coeffi\ we look at the second part of \dunkl. It is suggestive to
compare the matrix elements of $x(p,q)$ with the coordinates $x_i$ of the Dunkl
operator. In the Higgs sector the matrix field $X$ can be chosen to be
diagonal so that the identification relates the $x_i$ with the eigenvalues
of the matrix. But one has to take care for the possible breakdown of the 
$U(N)$ symmetry. The matrix $K_{ij}$ exchanges the modes $(n,m)$ in
\coeffi\ and commutes the spacial coordinates $(p,q)$. To
compensate the noncommuting part, one has to introduce two new indices
for the coordinates $(p,q)$ and an additional parameter, which leads to the
Macdonald polynomials, as already mentioned in Section 3. For this reason we
will assume that the matrix fields only depend on one spacial coordinate
$\theta$, parametrising the sphere $S^1$. With these simplifications the
creation operator $B^+_k$ takes the simple form of the Liouville action
\livs. To see this, identify the coordinates in \dunkl\ with the eigenfunction
of $S^1$. As stated in Section 3, the operator $B^+_k$ contains basically terms of the form 
$(D_i + \omega)x_i$. The first part then reduces to the derivative
$\d/\d\theta$, whereas the sum vanishes because of the cyclic structure of
$x_n$. But now the affine Lie algebra of $S^1$ is the Virasoro algebra, and
the shift operator, with its origin in matrix theory, can be reformulated in
creation modes $a_n$ of the minimal model \livs. 

To get an operator, independent of further indices, we make use of the
combination $\sum_n a_{-n}\LL_n$ and get the final form 
\eqn\scalarop{\DD = {1\over 2}\sum_{n}a_{-n}L_n -
             {1\over 2}\alpha_0 \sum_{n}(n+1)a_{-n}a_{n}\;.}
The advantage of this representation, instead of the more elementary form of
$\LL_n$, is its symmetry in the index $n$. In the case of commuting
variables, it is easy to give the general action on the string modes $a_n$.
The operator $\DD$ acts as a derivative, why we will call $\DD^k a_n$
the k-th derivative and denote it by $a^{(k)}_n$. For small $k$ we will also
make use of the notation $a'_n = \DD a_n$. The first derivative and its commutator
take the form 
\eqn\deriv{\eqalign{
a'_{n}\hskip 1em = & \;\; -nL_{n} +\alpha_0 n(n+1)a_{n}\cr
[a'_{n}, a_m] = &\; nm \left( a_{n+m} - \alpha_0 (n+1)\delta_{n+m}\right)\;.}}
The commutation relation generates a new algebra with many interesting
features. For example, in the case that the Liouville theory degenerates to
an ordinary conformal field theory at $\beta = 1$, the derivative reduces to the
operator $-n L_n$, and all possible states, generated by $\DD$ degenerate to
the classical vertex operator of string theory. Another interesting feature
is the degeneracy at $n=-1$. In Section 3 we observed a contradiction for
the case $Q_1=1$, where classically the untwisted as well as the twisted
conformal description are valid conformal field theory descriptions of the
moduli space. We see, that for any partition of $N = Q_5Q_1$ the
critical case of $n=-1$ is independent of the Liouville part of the action.
Therefore the operator mode $a'_{-1}$ can only shift the index by one, but
does not change the actual ground state. Although this does not solve 
the problem of the conformal charge and its
independence concerning the D1-brane charge, it circumvents the only
critical point in our discussion by eliminating this dependence for
$a_{-1}$. The other characteristics of the derivative $\DD$ are less
obvious. In anticipation of a discussion in the next section, we will show
here that the relations \deriv\ and thus the operator \scalarop\ are all we need
to calculate the Jack polynomials. 

In the previous sections we argued that the stack of $D5/D1$ branes only
allows $\Z_2$ twists for any number of $N>2$. If our arguments are correct,
this twist of the fields corresponds to the first derivative of the string
modes, which already determines the action of the operators $a^{(k)}_n$ on
the vacuum to be zero, because the ground state cannot be twisted. The
ansatz for the other state vectors, which will be motivated in the next
section, is the polynomial 
\eqn\asn{J_{(1^n)}(a,\beta=2)=
        {1\over n!}\left(a_{-1}+a'_{-1}\right)^n\,|0>\;.}
The single first derivative $a'_{-1}$ interweaves the D1-brane with the
remaining stack of D5-branes. Calculating the first three terms gives the
result 
\eqn\bsn{\eqalign{
J_{(1^1)}  = & \;a_{-1} \cr
J_{(1^2)}  = & \;{1\over 2}\left(a_{-2}+a_{-1}^2\right)\cr
J_{(1^3)}  = & \;{1\over 3}\left(a_{-3}+{3\over 2}a_{-2}a_{-1}+
               {1\over 2}a_{-1}^3\right)\;,}}
which reproduces the primary states \Astates\ up to a sign reversal of the
string modes. As expected, all these states belong to the Higgs sector and
after a convenient rescaling can be rewritten in the simple form of Schur
polynomials. The generating functions \reps\ now allows an even more
suggestive form as the vertex operator 
\eqn\vertexone{ e^{-\Phi(z)} =
                \exp{\left(-\sum_{n\in\Z}{(-1)^n\over n}q_n z^n\right)}\;,}
with the new bosonic operator modes $q_n$, obeying the commutation relation
\topcomm. This vertex operator can be understood as the effective field of
the $D5/D1$ brane system with the partition of D1-branes along the stack of
D5-branes, similar to an effective action in field theory. In the next section
we will be concerned with a better understanding of this construction in the
framework of moduli spaces \refs{\nakajimab, \lehn}.

For the above example the guess of a generating function has been quite simple,
but for more general cases one needs a strategy to solve the equation \asn.
In a first step one has to reformulate the defining equation in the Virasoro
modes as a recurrence relation in a polynomial ring. With the substitution
of $a_{-1}$ by the variable $x$, equation \asn\ becomes
\eqn\recrel{  A_n(x) = (x + x D) A_{n-1}(x)\;,\hskip 1em
              A_n(x) = \sum_{k=0}^{\infty}g[n,k]x^k }
with the action of $D$ determined by \deriv. Inserting the ansatz for
$A_n(x)$ one finally obtains a recurrence relation in the coefficients of
$g[n,k]$ 
\eqn\recrelg{  g[n,k]  =  g[n-1,k-1] + k\, g[n-1,k-1]\;. }
Which again is easily solved in the index $n$ by the generating function
$B_k(y) = \sum g[n,k] y^n$ and thus determines the coefficients $g[n,k]$. It
is interesting to be a bit more general and to modify the recurrence relation
\recrelg\ by an arbitrary integer $\alpha -1$
\eqn\recrelg{  g[n,k]  =  (\alpha -1)\,g[n-1,k-1] + k\, g[n-1,k-1]}
to obtain the equation
\eqn\recrelcoeff{\eqalign{
 \sum_{n} g[n,k] y^n = &\; (\alpha -1)y \sum_{n}g[n-1,k-1]y^{n-1}
                         +ky\sum_{n}g[n-1,k-1]y^{n-1}\cr
       B_k(y)         =&\; y(\alpha +k-1) B_{k-1}(y)\cr
                      =&\; y^{k}(\alpha +k-1)\ldots (\alpha-1)\;. }}
The only nonvanishing coefficient $g[k+1,k]$, inserted into the
generating function gives the result
\eqn\recfuncexp{\eqalign{
  V(z) =&\; \sum_{k=0}^{\infty} {B_k \over k!} z^k \cr
       =&\; \sum_{k=0}^{\infty}{(k+\alpha -1)!y^k \over (\alpha -1)! k!} z^k \cr
       =&\; {1\over (1-yz)^\alpha}\cr
       =&\; \exp{\left(-\alpha \sum_{n=0}^{\infty}{y^n\over n}z^n \right)}\;.}}
This derivation, being valid for arbitrary values of $\alpha$, it allows
in a convenient way to calculate the generating function in depenceny of the
parameter $\beta$.

All the primary states of the Higgs branch can be reproduced in the form of
\asn, but the construction of the nonprimary states is still an outstanding
problem. Following the same argument as above, that the $D5/D1$ brane system
demands first order derivatives only, the missing states are determined by 
\eqn\misstates{\eqalign{
J_{(1^3)}                    = & \;{1\over 3}a_{-3}+
                        {1\over 2}a_{-2}a_{-1}+{1\over 6}a_{-1}^3\cr
{1\over 1!} a_{-1}'J_{(1^2)} = & \;a_{-3} + a_{-2}a_{-1}\cr
{1\over 2!} a_{-2}'J_{(1^1)} = & \;a_{-3}\;,}}
which is in exact agreement with the orbit sum \summe\ after the
reversal of signs as above. This shows that all single particle as well as
multiparticle states have a representation as an polynomial in derivatives
of $a_n$. 

The restriction of our consideration to D1 and D5-branes made the
introduction of the first derivative necessary. But as has already been
noted in Section 3, the incorporation of D3-branes along the lines of
Section 2 force the introduction of even higher twists and thus higher
derivatives in the string modes. It is interesting to look at the
general structure, which the derivative $\DD$ generates by its iterative action on
the Liouville field $\phi$. The first three terms in this sequence, as
calculated from the differential operator \scalarop\ and the explicit form of
the Liouville action \liv\ are given by
\eqn\sequence{\eqalign{
V^{-1}=&\;\d\phi\cr
V^0=&\;{1\over 2}(\d\phi)^2-\alpha_0\d^2\phi\cr
V^1=&\;{1\over 3}(\d\phi)^3-\alpha_0\d\phi\d^2\phi+
                                  {1\over 3}\alpha_0^2\d^3\phi\;.}}
So each operation by $\DD$ increases the order of the partial derivative
$\d\phi$ by one. A generator $V^k$ of this sequence is therefore a bosonic
current of charge $s=k+2$. This is the main solution to the original
problem, namely to explain the origin of a CFT with conformal charge
$c=6N$ for any integer $N=Q_5Q_1$, although the original theory only
contained a massless field of at most spin two. As suggested by Vafa
\vafapuzz\ the elements of \sequence\ resemble the lowest currents of a
$W_{1+\infty}$ algebra, a symmetry valid not only for the twisted but
also for the untwisted sector, as learned from \commads. W algebras have many
interesting features of which we will review only some in the following. As
an introduction and for further references we recommend \refs{\viele, \pope}.

The general $AdS/CFT$ correspondence \largen\ supposes the duality
to be exact only in the large $N$ limit, what excludes all finite
dimensional W algebras. Furthermore, we will always need at least one spin-one
current, which gives a further restriction on $W_{1+\infty}$ algebras and
its tensor products with at most $W_\infty$. The bosonic realisation of the
$W_{1+\infty}$ algebra is sufficient for the description of the moduli space
of $K3$, but the introduction of $T^4$ already demands the incorporation of
the $N=2$ supersymmetric W algebra \viele, whose bosonic sector is
$W_{1+\infty}\times W_{1+\infty}$. The free field Lagrangian of the
underlying model is $L = \dbar\phibar\d\phi +\psibar\dbar\psi$ and respects
the necessary additivity of the number of cohomology classes and conformal
charges. As learned from \pope\ the analysis of the $W_{1+\infty}$ algebra is best
studied in the fermionic realisation 
\eqn\ferm{L={1\over 2}\d\psibar\psi-{1\over 2}\psibar\d\psi}
with central charge $c=1$. The resulting currents $V^k(z)$ form an
irreducible basis of the algebra regarding the spin, instead of the
multiparticle analysis from $AdS/CFT$. Of course this basis is not unique,
but a convenient choice is one for which the currents are quasiprimary
states with respect to the energy-momentum tensor. This allows to identify 
the W algebra currents with the single particle states of
the KK modes in a 1-parameter dependent basis
\eqn\basisa{V^i(z)=\sum_{j=0}^{i+1}\alpha_j(i,\lambda)\d^j\psibar\d^{i+1-j}\psi\;,}
with the coefficients 
\eqn\acoeff{\alpha_j(i;\lambda)={i+1\choose j}
           {(i+2\lambda+2-j)_j(2\lambda-i-1)_{i+1-j}\over(i+2)_{i+1}}\;,}
where the bracket $(a)_k$ is the ascending Pochhammer symbol for integer
$a$ defined by $(a)_k=(a+k-1)!/(a-1)!$. To compare this representation of
the currents with the higher derivatives of the bosonic modes of $\phi$, one
has to rebosonise the complex fermionic field $\psi$ the free scalar field
\eqn\subs{ \psi = e^\phi \hskip 2 em {\rm and} \hskip 2em \psibar = e^{-\phi}}
The bilinear terms $\d^j\psibar\d^{i}\psi$ in the sum \basisa\ now take the form 
\eqn\blabla{\d^j\psibar\d^{i}\psi  =\sum_{k=i+1}^{i+j+1}{1\over k}
           (-1)^{k-i-1}{j\choose k-i-1}\d^{i+j-k+1}P^{(k)}(z)\;,}
where the polynomials $P^{(k)}(z)$ are of Hermitian type:
\eqn\blebvl{P^{(k)}(z)=e^{-\phi(z)}\d^ke^{\phi(z)}\;.}
The expansion of \basisa\ with these redefinitions finally gives the currents
\sequence\ with the Liouville field as free bosonic particle \pope\ and parameter
$\lambda = -\alpha_0$ as improvement charge. The analog description for
$T^4$ with 8 even and 8 odd cohomology elements makes the introduction of
fermionic currents $G^i(z)$ and $\bar G^i(z)$ necessary \viele, and extends
the algebra to the $\NN=1$ supersymmetry $W_{1+\infty}(\lambda)$.
The $\lambda$ dependence makes the discussion rather complicated, why we
take the parameter to be zero. Then the additional currents take the form 
\eqn\addcurr{\eqalign{
G^i(z) = &\; \sum_{k=0}^i \gamma_k(i)\d^{i-k+1}\phibar\d^k\psi\cr
\bar G^i(z) = &\; \sum_{k=0}^i \gamma_k(i)\d^{i-k+1}\phi\d^k\psibar\;,}}
where the expansion coefficients $\gamma_k(i)$ are given by
\eqn\addcuupar{\gamma_k(i) = {(-1)^k\over 2^i}{(i+1)!\over (2i+1)!!}
                             {i\choose k}{i+1\choose k}\;.}
Actually it is not the field representation of the W algebra we will need,
but the algebra of modes and the mode depending conformal charge. The
final result for the $W_{1+\infty}$ algebra with $a$ and $\lambda$ set to
zero is then \pope
\eqn\completealg{[V_m^i,V_n^j]=\sum_{k\geq 0}
    q^{2k}\tilde{g}_{2k}^{ij}(m,n)V_{m+n}^{i+j-2k}+
    q^{2i}\tilde{c}_i(m)\delta^{ij}\delta_{m+n}\;,}
with the rescaling parameter $q$. For the coefficients $\tilde{g}_{2k}^{ij}(m,n)$
and further analysis of the $q$ dependence we refer to \pope. What we are
interested in is the mode dependence of the conformal charge $\tilde{c}_i(m)$
\eqn\confchone{\tilde{c}_i(m)= 
m(m^2-1)(m^2-4)\cdot\cdot\cdot(m^2-(i+1)^2)\tilde{c}_i}
with the coefficient
\eqn\confchtwo{\tilde{c}_i=
              {2^{2i-2}\left((i+1)!\right)^2\over (2i+1)!!(2i+3)!!}c\;,}
which shows that the conformal charge of the W algebra is already determined
by the conformal charge of Virasoro algebra and thus by the topology of the
underlying space. To be more precise, the charge $c$ in \confchtwo\ is the
second Chern class of the compact manifold $X$. For the interesting case of
$X=K3$ this is simply $c=24$, the conformal charge of the bosonic string.
The mode expansion of the most general $W_{1+\infty}(\lambda)$ algebra is
far more complicated, so that we shall only consider the differences to
\completealg.

For generic $\lambda$ this W algebra has a $\NN=2$ extension and not the
previously considered $\NN=1$ supersymmetry. But there is no contradiction,
because the algebra degenerates for $\lambda=1/4$ to $\NN=1$. The
understanding of this degeneracy offers a further connection between the
improvement charge $\alpha$, the parameter $\lambda$ and the supercharge of
the enveloping algebra ${\rm osp}(1,2)$ of $W_{1+\infty}(\lambda)$. The
commutator relation of the supercharges \viele
\eqn\supercharge{ [G_\alpha, G_\beta] =
                  \left( H+{1\over 2}\right)\epsilon_{\alpha \beta}\;,}
introduces the bosonic operator $H$, which commutes with any element
of the algebra. As noted in Section 3, the $AdS_3$ space has an
affine $SL(2,\R)\times SL(2,\R)$ symmetry \refs{\exclusion, \commads} with
second Casimir operator 
\eqn\casimir{\eqalign{
C_2 =&\; {1\over 2}\{L_1, L_{-1}\} - L_0^2 -
         {1\over 4}[G_{{1\over 2}}, G_{-{1\over 2}}]\cr
    =&\; {1\over 16}-{1\over 4}H^2\cr
    =&\; \lambda(\lambda+1/2)\;.}}
Introducing the Klein operator $K$, satisfying $K^2={\bf 1}$, this equation
can be solved for $H$
\eqn\klein{ H = 2\left( \lambda -{1\over 4}\right)K\;.}
The degeneration of the $\NN=2$ superalgebra at $\lambda=1/4$ to $\NN=1$ is
obvious from \supercharge\ and furthermore the value $\lambda=0$ is a
point of even higher degenerateness, as the complete fermionic part of the
algebra vanishes. 

The appearance of the Klein operator $K$ has an important impact on the
understanding of the W algebra from the mathematical point of view. 
Remember the D-brane charge dependence of the Liouville
charge $\alpha_0$ in terms of $\beta=Q_5/Q_1$. To compare our results with
the special case of moduli spaces with only one D1-brane inserted, let us
assume $\beta>>1$, which reduces \casimir\ to the simple expression
$H^2\sim Q_5/Q_1 {\bf 1}$, the volume of the surface of $X$. This relates
$K$ to its canonical class normalised by $K^2=1$. For $X=K3$ there is no
further condition on $\beta$, but the W algebra description of $X=T^4$ seems
only to be valid for $\lambda=1/4$ and fixes the relation between the
number of D5-brane charges and D1-branes.

Up to now we have shown that there are W algebras that reduce to
the expected conformal field theories in the limit $Q_1\to 1$. But as 
has been already noted in \commads\ the commutation relations of the vertex
operators in the large string limit \liv
\eqn\vertexop{ V_{jm\bar m}=\gamma^{j+m}\gammabar^{j+\bar m}
               \exp{\left( {2j\over \alpha_+}\phi\right)}}
is basically determined by the power of the field $\gamma(z)$ and its
derivatives. Setting the conformal charge to zero, i.e. $Q_5=0$,
reduces this system to the affine $S^1$ algebra $w_{1+\infty}$. In this
limit the difference between the insertion of $Q_1$ vertex operators and the
integration over the $Q_1$ covering of the complex plane $z\in\bf C$ vanishes.
This strongly suggests that the structure of the W algebra remains valid
also for $Q_5\neq 0$. In this case the leading commutator
contributions of the $w_{\infty+1}$ algebra for the expansion modes $v^i_n$
take the form \pope
\eqn\leading{[v^i_m,v^j_n] = 
\left((j+1)m-(i+1)n\right) v^{i+j}_{n+m}+
{c_0\over 12}m(m^2-1)\delta^{i,0}\delta^{j,0}\delta_{n+m} \;.}
The complete description of the algebra $W_{1+\infty}(\lambda)$ can be found
in \viele, but the structure of the leading terms is sufficient to calculate
the general form of vertex operators as the characteristic function of a
partition \lehn\ along the line of \vertexone. As a final step we are
able to show that the whole W algebra is generated by the operator
\scalarop. The Virasoro mode $a_n$ is represented by $v_n^{-1}$ in
the algebra \leading. Thus the first derivative $\DD a_n = a_{n}'$  can
be read off from \sequence\ as the commutator $[v_0^1, v_n^{-1}]=-2n v_n^0$,
which allows us to identify the scalar operator with the mode $\DD=-{1\over 2}v_0^1$ 
and the general form of the k-th derivative with 
\eqn\schrott{v_n^k = n^{-k}\DD^k\cdot a_n\;.}

\subsec{The Boundary of the Hilbert Scheme} 

In the Introduction we explained our understanding of D-brane dynamics as a
change of relative positions of branes in a $D5/D1$ stack. The description
of such a system depends on additional moduli parameters \tmod\ and \vmod,
which requires a more systematic analysis of 
the null states and its derivatives. If the
moduli space of a $D5/D1$ brane system would only depend on the product of
the brane charges $N=Q_5Q_1$, one simply had to sum over all possible
partitions, restricted by the number of marginal deformations in the twisted
sector. An explicit construction of the twisted vertex operators is
not required, but only the knowledge of their conformal weights,
reducing this calculation to a combinational problem in the $AdS/CFT$
framework. Thus we arrive at two problems. One is the introduction of
moduli parameters, which relate the cohomology classes of $X$ to the
topology of the Hilbert scheme $\MM$. The second and more complicated
problem is to find the generating function that reproduces the allowed
partitions of D-branes in the large $N$ limit of the $AdS/CFT$
correspondence. 

Let us concentrate on the first problem. Its solution already entered the
mathematical discussion in \refs{\nakajimab, \lehn}. Here we will motivate
their construction from a more physical point of view, which
already has been used in the construction of the vertex function \vertexone\
and the introduction of the twisted states as vectors in the space of
cohomology classes. The commutator relation for the bosonic modes \comm\
can be written in the symbolic form
\eqn\symbolic{ [\alpha^A_{n}(x),\alpha^B_{m}(y)]=n\delta^{AB}\delta_{n+m}
               <x,y> \hskip 2em {\rm for}\hskip 2em x,y \in K(X,\Z)}
with the scalar product \inter. This way the cohomology group is related
to the K-group $K(X,\Z)$ and the RR fields \ktheoryb\ pulled back from
the Hilbert scheme $X^{[N]}$. For a mathematical discussion we refer to
\lehn. From \ktheoryb\ and the
discussion of Section 4.2 we know that the intersection form \inter\
generates the N\'{e}ron-Severi lattice $NS(X)$. Now, the advantage of the
string theoretic description of \symbolic\ becomes apparent in the different
interpretations as RR fields, cohomology and K-theory. For a
stack of $D5/D1$ branes, for example, the result of the scalar product,
integrated over the space $X$ is simply $Q_5Q_1$, but the string modes
$\alpha_n, \psi_m$ in \comm\ have been defined after a rescaling by
$1/\sqrt{Q_5Q_1}$. In the following we will therefore understand the
product form $<x,y>$ as normalised by this factor $N$, which is the reason
why the moduli parameter $t$ \tmod\ does not enter the vertex function
\vertexone. But the choice $t=1$ fixes all
other parameters in the discussion. One example, already introduced in
Section 2 is the moduli of D3-branes, $v$, as in \vmod. This way the Virasoro
algebra becomes an algebra over the commutative ring of the
homology group of Hilbert schemes $X^{[n]}$ with $n = N + 1/2 Q_3^2$.

This lies at the core of the construction. Without the moduli parameters the
description of the Hilbert scheme by the Virasoro algebra would only be an,
although interesting, way to understand intersecting branes in two
dimensions. But now the 
information about the surrounding space is encoded in the moduli
parameters and the otherwise extremely complicated calculation of the
underlying physical information in terms of algebraic geometry reduces
to the simpler Virasoro algebra. In turn, the basic purpose of this
algebra is to determine how to multiply the moduli parameters. Still, it
takes the large $N$ limit or, to be more precise, the infinite sum over all
possible partitions to get the right picture. This again comes without
surprise, as string theory is determined not by an action like field
theory, but by the interactions, determined by vertex operators of the form
\vertexone. This hypothesis sheds a new light on many effects, typical for
string theory. As an example, the Virasoro modes of a D-brane are $a^i_n(v)$
as considered above. If our ideas are correct, the corresponding modes of an
anti D-brane are $\tilde{a}^i_{-n}(\bar v)$ with $\bar v$ the complex 
conjugate moduli parameter. The minus sign $-n$ takes
care of the opposite orientation of the anti D-brane compared to the
original D-brane. Classically, the parameter is one of the two complex
dimensions $u=x_6+ix_7$ or $v=x_8+i x_9$  with vanishing commutator. But this
changes if a selfdual NS B-field in the $(x_8,x_9)$ direction is turned on.
The N\'{e}ron-Severi
lattice $NS(X)$ depends only on the RR charges and does not change, but the
coordinates $(x_8,x_9)$ become noncommutative $[x_8, x_9] = i 1/2\theta$.
Let us abbreviate the commutator by a simple dot, then the two nontrivial 
relations are $v\cdot v = 0$ and $v\cdot \bar v = \theta$, with all
commutators with $u$ and $\bar u$ vanishing. The left and right modes of the
Virasoro algebra do not commute anymore, but take the relations
\eqn\relations{[a^i_n(x),\tilde{a}^j_m(\bar x)]=\theta\delta^{ij}\delta_{n-m}, 
\hskip 1em [L_n(x),\tilde{a}^i_m(\bar x)]=a^i_{-m+n}(\theta) \hskip 1em
                {\rm for} \hskip 1em x\in K(X)\;,}
with an analog algebra for the open string \ktheorya. The algebra differs in
two important points. First the Kronecker
delta $\delta_{n-m}$ guarantees the condition $N= \tilde N$ for the brane /
anti-brane partitions, necessary to fulfil the anomaly condition. The
second point is the effect of $L_n$ on $\tilde{a}^i_m$. It decreases the
number of antibranes by $n$. The missing factor of $-m$ guarantees that the
number of residual branes is identical to the number of modes. But the
algebra has another interesting interpretation. The moduli parameters
$(u,v)$ correspond to two-dimensional cycles in the space $X$. A D5-brane
and an anti D5-brane wrapping around the same cycle $\Sigma_1$, denoted by
$u$, anihilate into a D3-brane of the residual directions. Suppose, that a
second cycle $\Sigma_2$ exists with $\Sigma_1\cap\Sigma_2=1$ and
parametrised by $v$. It would then be quite natural to identify the
noncommuting parameter $\theta$ as an additional contribution 
to the second moduli parameter $v$. This
shows how the noncommutative geometry of the NS B-field and the lattice
structure of K-theory are related from the moduli space point of view.

Now that the incorporation of moduli parameters has been explained, we
can turn to the second problem of this section, the infinite sum over all
possible D-brane partitions. In general this is as simple as the
deformations of $X$ by moduli parameters, but in practice this is the
complicated part, and we have to refer to the article of M. Lehn \lehn\ for
a nontrivial example. In Section 4 we interpreted the DBI action of a brane
as the total Chern class of the endomorphismen bundle of the Hilbert scheme,
or to be more precise, as the Chern class of the tautological bundle
\lehn. We already gave the physical interpretation above, namely as the interaction
of a brane with a $D5/D1$ stack. In terms of the Virasoro modes $a_n(x)$ the
propagator $\chern^{-1}(u)$ acts on a single brane state by \lehn
\eqn\prevertex{\eqalign{
\CC(u)  = & \;\chern(u)\cdot a_{-1}(x)\cdot\chern^{-1}(u) \cr
 = & \sum_{n,k\geq 0}(-1)^n{\rk(u)-k\choose n} a_{-1}^{(n)}(\chern_k(u)x)\;,}}
with $u\in K(X)$ and $x\in H^*(X,\Z)$. The binomial coefficient is simply
obtained by a little combinatoric and the restriction that the n-fold
product of a $d$ dimensional manifold only allows a Chern class of order
$nd$ and partitions thereof. The rank of the K-group element $u$ enters
the expansion, which has a simple interpretation in string theory as the
rank of the difference gauge bundle $(E,F)$ of the branes and antibranes
$\rk(u)=\rk(E)-\rk(F)$. For completeness we introduce the Chern
character of the tautological bundle of the Hilbert scheme, since it
has an intuitive interpretation and completes the argument that justified
the representation of the higher twist modes as derivatives of the primary
states of the Higgs branch \misstates. The derivative \scalarop\ is of order
one in the Chern classes and from the point of field theory its action can
be understood as the insertion operator of the first Chern class
$\chern_1(\FF)$. The formal Chern character $\exp{(\FF)}$ defines the
expansion coefficients for the generating operator $\DD$
\eqn\character{\eqalign{
e^{\DD}a_{-1}(\ch(u)x)=&\;\ch(u)\cdot a_{-1}(x)\cr
= &\sum_{n,k\geq 0}{(-1)^n\over n!}a_{-1}^{(n)}(\ch_{k-n}(u)x)\;.}}
The advantage of the Chern characters is the homogeneity of the operator
expansion in comparison to the Chern classes, who interchange the degrees
of the derivative $\DD$ and the modes $a_n$. For the special case of the
affine $S^1$ algebra, introduced in the previous subsection, the action of
the Chern character onto a special partition can be calculated by
combinatorics \lehn. Take the two partitions
$\lam=(\lam_1,0,\ldots,0,\lam_n,\lam_{n+1},\ldots)$ and
$\lam'=(\lam_1+n,0,\ldots,\lam_n-1,\lam_{n+1},\ldots)$ both of the same
degree and note by $a_\lam$ the corresponding symmetric
function. As shown in \lehn\ the Chern character $\ch_{n-1}$ relates $\lam'$
with partitions of higher length
\eqn\length{\ch_{n-1}a_{\lam'}={\lam_1+n\choose n}a_\lam+\ldots\;,}
where we again redefined all modes $a_n$ by a minus sign. This finally
justifies the identification of the derivative as the twist operator
\misstates. In \lehn\ M. Lehn comes up with another interpretation of $\DD$,
more appropriate from a mathematical point of view. Let $X^{[1,1,\ldots]}$ be
a Hilbert scheme of order $N$, then the action of $\DD$ maps it to
$X^{[2,1,\ldots]}$ the ``boundary'' of the Hilbert scheme. 

Finally we are able to define the vertex operator in full generality with
consideration of additional moduli parameters. As long as one is operating
with BPS states, the effective field along example \vertexone\ is a
holomorphic function of the complex coordinate $z\in\IC$ only, defined by
the interaction along two dimensions. 
\eqn\finalresult{\eqalign{
e^{\Phi(z)} = &\;\exp{\left( \int_{X}\CC(u)z \right) } \cr
= &\sum_{n=0}^{\infty}\int_{X^{[n]}}\left({1\over n!}\CC^n(u)\right) z^n\;,}}
with the definition of $\CC(u)$ from \prevertex. Here we have to be more
specific with regard to the integration over the Hilbert scheme $X^{[n]}$.
As mentioned above, the interpretation of the Hilbert scheme as the symmetric
product of $X$ is defined by a sum over all partitions. The integration over
$X^{[n]}$ now compares a specific partition with the one of $\CC^n(u)$. It is
 quite natural to compare the Hilbert scheme, deformed by the blow
ups along the singular points, with the trivial symmetric product of $n$
copies of $X$. In the representation of the bosonic Virasoro algebra this
is simply the mode $\CC(1_X) = a_1$, with no further moduli inserted. The
effective vertex operator of this special partition takes the simple form 
\eqn\vertexnull{e^{\Phi_1(z)} = \;e^{a_1z}\cdot 1_X\;.}
This is one example, where the naive correspondence between vertex
operators of string theory differs from the one of the Hilbert scheme. But
there is one more important case. To define the vacuum structure
of the underlying D-brane configuration, one still has to determine the null
mode of the field $a_0$, which is defined by the relation
$a_0|\chi_{0,0}> = \alpha_0|\chi_{0,0}>$. The vacuum state connects the effective 
field picture with the topological information of $X$
encoded in the conformal charge of the algebra. The third operator which is
important for our discussion and characterises the Hilbert scheme $X^{[0]}$
is therefore
\eqn\vertexvacuum{e^{\Phi_0(z)} = \;e^{a_0 z}\;.}
For all these vertex operators the incorporation of antibranes is obvious
in the commutative case, as the two sectors are independent. The only
crucial condition one has to consider is the restriction from
K-theory for D-brane / anti D-brane interactions \ktheorya.

Up to now, we only considered the effect of interacting D-branes along two
dimensions and the consequences for the residual compact space. But what
about the noncompact directions as occurring in the algebra \algb? The
N\'{e}ron-Severi lattice $\NS(X)$ defines the signature of the Hilbert
scheme, whereas one must not forget the original metric of the
ten-dimensional spacetime. How do the two metrics combine? For example,
consider the vertex operator 
\eqn\try{ V_{-{1\over 2}, -{1\over 2}} =
         e^{-{1\over 2}\phi(z)}e^{-{1\over 2}\tilde\phi(\zbar)}
	 k\cdot\Gamma_{\alpha\beta}S^\alpha S^\beta e^{ikX(z,\zbar)}\;,}
with $\alpha$ the spinor index for the non compact directions and $\beta$ the
index from the $\NS(X)$ lattice. The problem is best studied for the Clifford algebra,
which only depends on the signature and the dimension of a manifold. For
simplicity we choose $X = K3$ with $\rk(\NS(K3))=24$. Modular invariance of
the partition function forces a space of 10 or 26 dimensions, which leaves
only one possibility. Two of the 10 dimensions are fixed by the directions
of the D-brane interactions, so that the residual 16 dimensions have to
be ``compactified'', analogous to the heterotic string. But this formal
compactification has the consequence of changing the Hilbert scheme, without
necessarily affecting the topology of the original manifold $X$. The only free field,
not determined by the cohomology of $X$, is the NS B-field. From the point of
the effective vertex operator \finalresult, this additional twist can be
interpreted as a torsion element in the $\NS(X)$ group; but we have no
example for such a twist. The second condition to take care of, is the
signature of the metric. Before compactifying 16 of the 24 dimensions of the
$K3$ lattice, one has to determine the number of timelike directions. It is
not necessary to have exactly one time direction and string theories with
different signatures have been analysed \hull, but the energy of such a
theory may be indefinite and thus the moduli space of this manifold need not to
be stable. We will simply assume, that the signature of
the NS lattice is of the form $(1, D-1)$ to get a stable moduli space and a
closed vertex operator algebra.

Having explained the most basic ideas of the construction, we
will show how to apply these methods to string theory. The simplest
possible generalisation of the vertex function of the $D5/D1$ system
\vertexone\ is the embedding of D3-branes. For the manifold $X$ this is
equivalent to the embedding of a complex curve $\Sigma$ and thus the
introduction of a holomorphic line bundle $-\OO(H)$ of first Chern class
$\chern_1=H$. With the introduction of the moduli parameter \vmod\ a
calculation similar to \vertexone\ becomes impracticable. We will
not try to gain information from the effective vertex operator or the
resulting effective Virasoro algebra, but determine the 3-point correlator
instead
\eqn\threepoint{\eqalign{
<e^{\Phi_0(z=0)}\cdot e^{\Phi_1(z=1)}\cdot e^{\Phi(z)}> = & 
<a_0\cdot \left(\sum_{m=0}^\infty {1\over m!}a_1^m\right)\cdot 
\left( \sum_{n=0}^{\infty}\int_{X^{[n]}}\left({1\over n!}\CC^n(u)\right)
z^n \right)>\cr
= & c_0\sum_{n=0}^{\infty}{1\over n!}<{1\over n!}a_1^n| \CC^n(u)> z^n\cr
= & c_0\sum_{n=0}^{\infty}N_n z^n\;.}}
There are two possibilities for calculating the tree graph. The first
one is a generalisation of the operator product expansion to the W
algebra. But unfortunately not all coefficients are reproduced correctly,
although the basic structure of the exact solution is obtained. The
alternative is based on the step-by-step calculation of the coefficients
$N_n$. After integrating over the trivial Hilbert scheme, the connection between
the moduli parameter $v$ and the Chern classes has still to be determined. For the
simple case considered here, the embedding of the line bundle $-\OO(H)$ into
$X$ identifies the parameter as the Chern class
\eqn\exmod{v=\chern(-\OO_X(H))={\chern(\OO_X)\over\chern(-\OO_H)}=1-H+H^2\;.}
In addition to the rank $\rk(-\OO(H))=-1$ of the bundle, this has to be
inserted into the formula for the Chern classes of the Hilbert scheme
\eqn\classe{\CC(-\OO(H)) = \sum_{n=0}^\infty a_{-1}^{(n)}v^{n+1}\;.}
The exact calculation with the mathematical interpretation of the individual
coefficients $N_n$ has been done in \lehn. Here we will only quote the
result. After the coordinate transformation 
\eqn\changecoord{ z = {k(1-k)(1-2k)^4\over (1-6k+6k^2)^3}\;,}
the generating function of the 3-point function \threepoint\ becomes 
\eqn\genfunction{\sum_{n=0}^{\infty}N_n z^n=
                {(1-k)^a (1-2k)^b\over (1-6k+6k^2)^c}\;,}
with the abbreviations $a=HK-2K^2$, $b=(H-K)^2+3\chi(\OO_X)$ and
$c={1\over 2}H(H-K)+\chi(\OO_X)$ with the canonical bundle $K$ and the Euler
characteristic $\chi$. The last of the three coefficients is the Euler
characteristic $\chi(\OO_H)$ as determined by the adjunction formula
\zaslovyau. 

In the framework of W algebra the calculation of the coefficients $N_n$
is not appropriate for the OPE. Instead one should consider the generating 
function \lehn 
\eqn\genalter{\sum_{n\leq 0}N_n z^n=\exp{\left(-\sum_{m>0}
              {(-1)^m\over m}d_m z^m\right)}\;,}
with the coefficients $d_m$ as calculated in \lehn. For $m>1$ these are
linear combinations of the topological values $H^2$, $HK$, $K^2$ and
$\chi(X)$. And although we have been able to reproduce the structure of the
coefficients, we failed to get the correct factors by the W algebra. Thus
what is still missing, is the correct incorporation of the ghost action to get
a manifest BRST invariant formulation. A guiding hint shows up from the
additional condition on the canonical class $K$, as for dimensional reasons
it has to obey $K^3=0$. But we believe that even then the
sum \genalter\ has to be regularised, as the spin dependent conformal charge
$C_{00}(s)=-(6s^2-6s+1)$ is divergent. This missing calculation scheme prevents us
from a better understanding of the D-brane interactions and the difference
between $Q_1=1$ and $Q_1>1$. But nonetheless, the generalisation of the
$AdS/CFT$ to the W algebra is an important step in the understanding of
M theory for finite brane configurations $N$.

\subsec{Generalisation to Six Dimensions}

During the entire discussion of the generalised $AdS/CFT$ correspondence we
assumed the supersymmetric background to be a four-dimensional Calabi-Yau
manifold and the D-branes interacting along two dimensions. But this is of
course neither the only possible background nor the only possible way of
D-brane interactions, and at the end of this article we will give an outlook
to higher dimensional compactifications. Incidentally, this generalisation is
necessary for the type I string, since the fusion of D-branes / anti
D-branes takes place along four dimensions \ktheorya\ instead of the
considered two. In principle, the construction carries over to any complex
manifold with hyper-K\"{a}hler structure and any number of moduli
parameters. The four-dimensional spaces considered above allow one
parameter only, as the submanifold itself has to be even dimensional. The
construction of the previous subsection is therefore of quite general
value. As an example, take a six-dimensional Calabi-Yau $X$. Again we start
with a stack of black D1-branes and their dual D5-branes, wrapped around $K3$. 
For the $S^3$ of the previous
$AdS_3\times S^3\times K3$ background we choose the fibration
$S^1\times S^2$ by wrapping D3-branes around $S^2$. A further twist finally
maps $S^2$ to $T^2$. The D-brane configuration is still invariant
under T-duality along the four dimensions of $K3$ and the twisted $T^2$.
Wrapping six directions of a D7-brane around $T^2\times K3$ determines a
further twist of the product manifold to get the Calabi-Yau threefold as an
elliptic fibration of $K3$. After this second twist the original charge
duality of the $D5/D1$ system, which was important in our analysis of the
$AdS/CFT$ correspondence, gets lost. Which shows that it is not possible to extend
the previous construction to higher dimensions with only one Liouville
field. Instead we take two copies of the $AdS_3\times S^1\times T^2\times K3$
background and introduce two Liouville fields $(\phi_1, \phi_2)$ with the
moduli $(t_1, t_2)$ parametrising the volume of $T^2$ and $K3$,
respectively. With the volume of the Calabi-Yau set be one, the
twist between the two manifolds is determined by the linear combination
$1=q t_1+p t_2$ for $p,q\in\Z$. Interactions of the two fields have a
classical formulation by a $W_3$ algebra with conformal charge
$c=q c_1+p c_2$. Further information about the elliptic fibration is
encoded in the anomaly term, but we will not enter this discussion here.


\newsec{Conclusions}
In this article we dealt with the construction of an infinite Hilbert scheme on a
compact manifold and in this way generalised the $AdS/CFT$ correspondence as
introduced by Maldacena \largen. Although the question of the
missing states found a satisfactory solution here, many
aspects have to be left to future investigations. In the Introduction we
explained that the only degrees of freedom are the motions of branes in the
supergravity background. Because we restricted the discussion to the $X$ part of 
$AdS_3\times S^3\times X$, the effects of the flat directions to the Hilbert 
scheme did not contribute and thus a better understanding of the
dynamics of branes in this framework is still missing. But before these
additional dimensions can be incorporated into the W algebras, better
calculation technics have to be developed.

One further point, we think is important for a better understanding, is the puzzle
of the D1-brane charge $Q_1$. Maybe, there is a deeper connection between
the moduli space of Enriques surfaces and the supersymmetric string, as
their torsion dependence allows more general ``effective'' fields as the
moduli spaces of Calabi-Yau's. An aspect connected to this is the
generalisation of the DBI action. In Section 5 we introduced an iterative
contruction, as the number $N$ of D-branes had to vary. But for finite $N$
it should be possible to find a formulation in the gauge fields. For the
$D5/D1$ brane system considered above, the intregral over the surface of the
Hilbert scheme is classically known to be $\beta = Q_5 / Q_1$, or from the
point of K-theory, this is the difference between the bundle of D5-branes
and D1-branes. But we have not been able to identify the additional
contributions in string theory to justify our assumption. 

Another problem concerns the space $\MM$. In principle,
it contains all parameters of the moduli space of M theory and so should be
related to all other little string theories. This does not seem to be the case
in general. As motivated by mirror symmetry, our construction is general
enough to relate type IIA to IIB string theory, but we failed in the cases of
the heterotic and open strings. Because the moduli space $\MM$ depends
basically on the topology of $X$, only terms which do not generate a change
of the topology may be added to \coproduct. One natural generalisation
is therefore the incorporation of an affine space $\AA$. It does not contribute to
the topology of $X$, but provides enough space to take care of additional
twists in the moduli space. As generalisation of the construction \coproduct\ 
for the heterotic and open string theories we suggest
\eqn\summed{\tilde{\MM} =
\sum_{N=1}^{\infty}\;\coprod_{\nu:\;\rm{partition\;of\;N}}[X]^{\nu}\times
[\AA]^{N-|\nu|}\;.}
What is the interpretation of $\AA$ in physical terms? From F theory \ftheory\
we know that IIB compactified on an elliptically fibered $K3$ is dual to the
heterotic theory on $T^2$ with the moduli of $K3$ encoded in the
prepotential. But duality via M theory suggests first a resolution of
the singularities on $K3$ before the orbifold construction can be
performed. Only then can the resolved space be blown down to a simple $T^2$,
if possible at all. Now suppose that the $K3$ depends on $N$ moduli. The
most general prepotential is a linear combination of all possible
resolutions of $K3$ after the orbifolding, i.e. a formal polynomial 
\eqn\fomalresol{ a_0 X^{[0]} + a_1 X^{[1]} + \ldots a_N X^{[N]}
               \hskip 1em {\rm with }\hskip 1em  \#\, a_k \leq {N\choose k}\;,}
of degree $N$ and at most $2^N$ coefficients, corresponding to all possible
combinations of the $N$ linear independent sections in $K3$. The space $\AA$
is then generated by the functions $a_0, \ldots a_N$, depending on the
residual moduli parameters, not contained in the corresponding symmetric sum
$X^{[k]}$. But most elements in $\AA$ are zero because of the modular
invariance of the elliptic fibres, so that in the case of an irreducible
$K3$ the polynomial reduces to $j_N X^{[0]} + X^{[N]}$ with the $j$-function
of the elliptic fibres of $K3$.  Of course, the other direction of the
duality has to work the same way. Consider the heterotic string on a
singular $T^2$. The dual F theoretic description has to be on an
elliptically fibred $K3$ with moduli parameters generated during
the blowing up of the singularities. The prepotential of IIB thus depends on
a polynomial  
\eqn\fomalresol{b_0\tilde X^{[0]}+b_1\tilde X^{[1]}+\ldots b_N\tilde X^{[N]}\;,}
with functions in the dual space $\BB$. What is interesting at the
coefficients $a_k$ and $b_k$ is their similarity to cohomology elements.
Furthermore, it would be interesting to study the D-brane spectrum of the
Hilbert schemes $X^{[k]}$ along \douglas\ to understand the brane
interactions analogously to the $D5/D1$ systems studied in this article.
\bigskip
\centerline{\bf Acknowledgements}

This work is supported by the DFG-Graduiertenkolleg ``Analyse und Konstruktion in
der Mathematik''.

\listrefs
\end